\documentclass[journal,transmag]{IEEEtran}
\IEEEoverridecommandlockouts

\usepackage{amsmath}
\usepackage{amssymb}
\usepackage{mathrsfs}
\usepackage{amsfonts}
\usepackage[linesnumbered, ruled]{algorithm2e}
\usepackage{setspace}
\usepackage{cite}
\usepackage{multirow}
\usepackage{color}
\usepackage{tabularx}
\usepackage{amsthm}
\usepackage{algorithmic}
\usepackage{graphicx}
\usepackage{multicol}
\usepackage{subfigure}
\usepackage{bm}
\usepackage{booktabs}

\begin{document}

\setlength{\textfloatsep}{0\baselineskip plus 0.2\baselineskip minus 0.2\baselineskip}

\title{Joint Power and User Grouping Optimization in Cell-Free Massive MIMO Systems}

\author{\IEEEauthorblockN{Fengqian Guo, Hancheng Lu and Zhuojia Gu}

\thanks{\IEEEcompsocthanksitem
This work was supported by National Science Foundation of China under Grant 61771445, 61631017, 91538203.}

\thanks{F. Guo, H. Lu and Z. Gu are with CAS Key Laboratory of Wireless-Optical Communications, University of Science and Technology of China, Hefei 230027, China. (Email: fqguo@mail.ustc.edu.cn, hclu@ustc.edu.cn, guzj@mail.ustc.edu.cn).
}
}

\maketitle
\begin{abstract}

To relieve the stress on channel estimation and decoding complexity in cell-free massive multiple-input multiple-output (MIMO) systems, user grouping problem is investigated in this paper, where access points (APs) based on time-division duplex (TDD) are considered to serve users on different time resources and the same frequency resource. In addition, when quality of service (QoS) requirements are considered, widely-used max-min power control is no longer applicable. We derive the minimum power constraints under diverse QoS requirements considering user grouping. Based on the analysis, we formulate the joint power and user grouping problem under QoS constraints, aiming at minimizing the total transmit power. A generalized benders decomposition (GBD) based algorithm is proposed, where the primal problem and master problem are solved iteratively to approach the optimal solution. Simulation results demonstrate that by user grouping, the number of users served in cell-free MIMO systems can be as much as the number of APs without increasing the complexity of channel estimation and decoding. Furthermore, with the proposed user grouping strategy, the power consumption can be reduced by 2-3 dB compared with the reference user grouping strategy{, and by 7 dB compared with the total transmit power without grouping.}

\end{abstract}


\IEEEdisplaynontitleabstractindextext
\IEEEpeerreviewmaketitle

\section{Introduction}\label{Introduction}
\newtheorem{def1}{\bf Definition}
\newtheorem{thm1}{\bf Theorem}
\newtheorem{lem1}{\bf Lemma}
\newtheorem{cor1}{\bf Corollary}
\newtheorem{pro1}{\bf Proposition}

The rapid growth of mobile traffic, especially high volume video traffic, leads to pressing need for high throughput in mobile networks\cite{9020157}. To cope with such situation, massive multiple-input multiple-output (MIMO) emerges as a promising technique \cite{7835110,7886292, Ngo2015,7031971}. In massive MIMO, massive antenna arrays are deployed to simultaneously serve many users on the same time-frequency resource, with which high spectral efficiency is achieved. By distributing numerous antennas in a wide area, the concept of cell-free massive MIMO \cite{Ngo2015,Ngo2017} has been proposed recently and attracted much attention from academic and industrial researchers. Essentially, cell free massive MIMO is an integration of massive MIMO and distributed MIMO, which is expected to exploit benefits of these two techniques. In cell free massive MIMO, many geographically located access points are equipped with single or a few antennas. They serve a much smaller number of users coherently on the same time-frequency resource, ensuring uniformly good quality of service (QoS) for all users. Consequently, cell boundaries are eliminated. Moreover, a central processing unit (CPU) is introduced to coordinate data transmission at different APs through high-capacity backhaul links connecting these APs. Compared with small-cell systems, existing studies have shown that cell-free massive MIMO systems can significantly improve per-user throughput. However, at the cost of much more backhaul overheads\cite{Ngo2015}.

Many research attempts have been done to improve the performance of the cell-free massive MIMO systems. Among them, power control has been addressed, which is globally optimized by CPU to realize uniformly good services for all users in a wide area. The pioneer work on cell-free massive MIMO was done in \cite{Ngo2015}, where max-min power control is performed to maximize the lowest user throughput. After that, the max-min power control problem is investigated under various scenarios\cite{Ngo2015,Nayebi2017,Zhang2018,Bashar2019}. For conjugate beamforming and zero-forcing (ZF) precoding, low complexity power control algorithms based on the max-min criterion were developed in \cite{Nayebi2017}. In \cite{Zhang2018}, a max-min power control algorithm was proposed with consideration of transceiver hardware impairments. The authors in \cite{Bashar2019} studied the uplink max-min signal-to-interference-plus-noise ratio problem and obtained a globally optimum solution with an iterative algorithm. In the downlink cell-free massive MIMO systems, power control can be optimized to maximize the energy efficiency \cite{Ngo2017a}. Furthermore, power control has also been jointly considered with load balancing \cite{Nguyen2018}, backhaul  \cite{8422865}, fronthaul \cite{Femenias2019}, etc.

There still remain some deficiencies in research on cell-free massive MIMO. In order to implement beamforming in a hardware-friendly way or to eliminate co-channel interference by zero-forcing precoding, in general, the number of antennas in massive MIMO systems is assumed to be significantly larger than the number of users \cite{Chen2017,Lu2014}. Similarly, in cell-free massive MIMO systems, since each AP is assumed to be equipped with one or a few antennas, the number of served users is much smaller than the number of APs \cite{Chen2018,Ngo2017}. To serve more users, much more APs should be deployed. Correspondingly, the hardware cost and system complexity will be significantly increased \cite{Nayebi2017}\cite{Zhang2018}. Furthermore, to ensure the accuracy of channel estimation, length of pilot sequence is usually assumed no less than the number of users\cite{Ngo2017a, Li2018}. However, the number of samples in each coherence interval are limited. Hence, the length of data in each coherence interval will reduce as the number of users increase. Additionally, users have diverse QoS requirements. Requirement satisfaction is more important for users than fairness. In this case, widely-used max-min power control is no longer applicable.

To address these issues, in this paper, we investigate the time-division duplex (TDD) based cell-free massive MIMO systems\cite{Femenias2019}. Users are divided into different groups according to their assigned time-slots, then channel estimation and decoding are applied within each group. By doing so, both pilot overheads and system complexity can be significantly reduced. In addition, we perform power allocation to satisfy the QoS requirements of users, instead of max-min power control. The main contributions are described as follows.

\begin{itemize}
\item {In the downlink TDD based cell-free massive MIMO systems considering user grouping, we first introduce and analyze the main processes of uplink training and downlink payload data transmission. Then we derive the minimum power constraints under diverse QoS requirements after user grouping. Based on our analytical work and the transmission process with user grouping, we formulate the joint power allocation and user grouping problem with both conjugate and ZF beamforming under user QoS constraints, with the goal to minimize the total transmit power.}

\item We convert the problem into a form that can be handled by generalized benders decomposition (GBD) method. With GBD, we first decompose the problem into the primal problem (i.e., power allocation problem) and master problem (i.e., user grouping problem). Particularly, the user grouping problem is relaxed and then the relaxed problem is converted into a problem of searching for some special negative loops in a graph composed of users.

\item Based on the GBD method, we propose an iterative algorithm{ , which is feasible for both conjugate and ZF beamforming,} to approach the optimal solution to the converted joint power allocation and user grouping problem. In each iteration, the upper bound and lower bounds are obtained by solving the primal problem and the master problem, respectively. The gap between these two bounds is reduced iteratively. Therefore, the proposed iterative algorithm is provably convergent. Furthermore, to solve the master problem within polynomial time, a fast greedy suboptimal algorithm is proposed.

\end{itemize}

Simulation results validate the convergence and optimality of the proposed algorithms, and demonstrate that by user grouping, the number of users served in cell-free MIMO systems can be as much as the number of APs without increasing the complexity of channel estimation and decoding. Furthermore, with the proposed user grouping strategy, the power consumption can be reduced by 2-3 dB compared with the random user grouping strategy{, and by 7 dB compared with the total transmit power without grouping.}

The rest of this paper is organized as follows. In Section II, we give the model of the downlink cell-free massive MIMO systems, and formulate the joint optimization problem of power allocation and user grouping to minimize the total transmit power. In Section III, we decompose the problem into a power allocation problem and a user grouping problem. Problem analysis and solutions are also described. In Section IV, we relax and solve the master problem based on graph theory.  The system performance is evaluated in Section V. Finally, we give the conclusion in Section VI.

\begin{table}
\setlength{\abovecaptionskip}{-0.04cm}
\setlength\abovedisplayskip{1pt}
\setlength\belowdisplayskip{1pt}
\fontsize{8}{12}\selectfont
	\centering
	\caption{Main Notations}
	\begin{tabular} {|m{30pt}|	m{180pt}|}
    \hline
		Symbol		& Description							 	\\
    \hline
		$ G $ 		& Number of groups  										\\	
    \hline
		$ N $ 		& Number of users 									\\	
    \hline
		$ M $ 		& Number of AP 									\\		
    \hline
		 $p_{mn} $ 			& Power control coefficient of the transmit power that AP $m$ allocated to user $ n $  							\\
    \hline
 $q_{mn} $          & $\sqrt{p_{mn}}$ 	\\
    \hline
		$ \gamma_n $ 				& Target SINR of user $ n $	\\
    \hline
		$ \bm{x} $ 				& User grouping matrix 				\\
    \hline
		$ P_t $ 				& Total transmit power		\\
    \hline
		$\beta_{mn} $          & large-scale fading between AP $m$ and user $n$					\\	
    \hline
	\end{tabular}
	\label{table1}
\end{table}

\emph{Notations:} Vectors and sets are denoted by bold letters. $\lceil \cdot\rceil $ denotes the ceiling function. $\mathbf{A}^H$ and $\mathbf{A}^*$ denote the conjugate transpose and conjugate of $\mathbf{A}$, respectively.

\section{Systems Model and Problem Formulation} \label{sec:02}

We consider a downlink TDD based cell-free massive MIMO system where $M$ single-antenna APs and $N$ single-antenna users are randomly located in a wide area as shown in Fig. \ref{fig:01}. In the traditional cell-free massive MIMO system where all users sharing all coherence intervals or time-slots as shown in Fig. \ref{fig:01}(a). To relieve the stress on channel estimation and decoding complexity, we divide users in groups according to their assigned time-slots. Users assigned the same time-slot form a group. The time-slots assigned to different groups are assumed to be orthogonal.  { In this paper, we assume that the number of users is greater than the number of groups.} In Fig. \ref{fig:01}(b), users are grouped into 3 groups and served on 3 orthogonal time-slots, respectively. Both channel estimation and decoding are performed within the group. {There are two types of training, i.e., large-scale training and uplink training. The result of large-scale training is assumed to be accurate. The interval between two times of large-scale training is named as $\tau_{Lc}$, and {the} interval between two times of uplink training is named as $\tau_{c}$.} Each $\tau_{Lc}$ is composed of one large-scale training phase and some time-slots, and each time-slot contains one coherence {interval}. In general, we assume that $\tau_{Lc}\gg \tau_{c}$ \cite{Femenias2019}.

{
\setlength{\abovecaptionskip}{-0.02cm}
\setlength\abovedisplayskip{-0.02cm}
\setlength\belowdisplayskip{-1cm}
\begin{figure*}[htbp]
	\centering
	\includegraphics[scale=0.18]{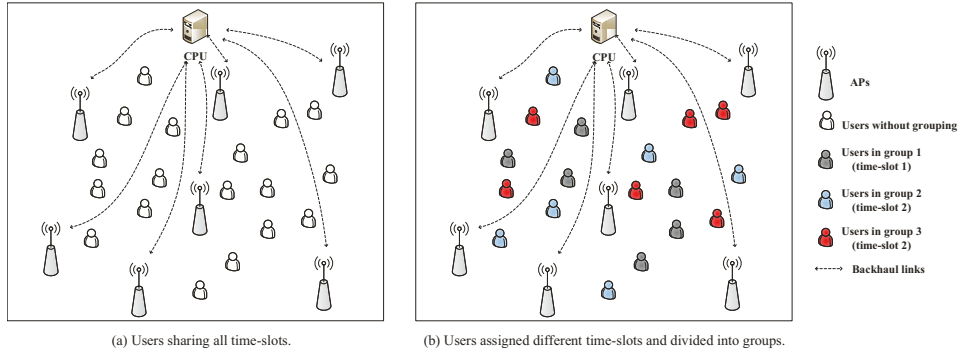}
	\caption{Illustration of Downlink Cell-Free Massive MIMO System.}
	\label{fig:01}
\end{figure*}}

The channel between AP $m$ and user $k$ on time-slot $g$ is modeled as {$
h_{gmn}=\sqrt{\beta_{mn}}\varsigma_{gmn},~~~     1\leq m\leq M, 1\leq g\leq G, 1 \leq n\leq N$, where $\beta_{mn}$ and ${\varsigma}_{gmn}\sim \mathcal{C}\mathcal{N}(0,1)$  denote the large-scale fading and small-scale fading between AP $m$ and user $n$ on time-slot $g$, respectively. In the remainder of this paper, we assume that large-scale fading $\beta_{mn}$ is known to all APs and users.
{Considering that the channels in different time-slots do not completely independent with each other, in this paper, the large-scale fading is assumed to remain constant across all time-slots between two large-scale training.} Some major notations are listed in Table I.

After user grouping, cell-free massive MIMO transmission in each group within a coherence interval consists of three phases: uplink training, uplink payload data transmission and downlink payload data transmission as shown in Fig. \ref{fig:01}. In this paper, we focus on the joint optimization of user grouping and power allocation in downlink cell-free massive MIMO system. Therefore, uplink training phase and downlink payload data transmission phase are introduced as follows.

\subsection{Uplink Training} \label{sec:02A}
{
\begin{figure*}[ht]
 	\centering
 \begin{equation}\label{0001}
\setlength\abovedisplayskip{0pt}
\setlength\belowdisplayskip{0pt}
\begin{aligned} &{\mathrm{SINR}_{n}^{MRT}}=\tfrac{\mathbb{E}[|Y_{us}|^2]}{\sigma^2+\mathbb{E}[|Y_{in}|^2+|Y_{ce}|^2+|Y_{lc}|^2]}=\tfrac{\mathbb{E}[|Y_{us}|^2]}{\sigma^2+\mathbb{E}[|Y_{in}|^2]+\mathbb{E}[|Y_{ce}|^2]+\mathbb{E}[|Y_{lc}|^2]}\\
=&\tfrac{\left({\sum\limits_{^{m=1}}^{M}\sqrt{p_{mn}}\sum\limits_{g=1}^{G}x_{gn}\alpha_{gmn}}\right)^{2}}
{\sigma^2+\sum\limits_{_{g=1}}^{G}x_{gn}\sum\limits _{^{i=1}_{i\neq n}}^{N}x_{gi}\sum\limits_{_{m=1}}^{M} {p_{mi}}\alpha_{gmn}\alpha_{gmi}
+\sum\limits_{_{g=1}}^{G}x_{gn}\sum\limits _{_{i=1}}^{N}x_{gi}\sum\limits_{_{m=1}}^{M} {p_{mi}}(\beta_{mn}-\alpha_{gmn})\alpha_{gmi}+\sum\limits_{_{g=1}}^{G}x_{gn} \sum\limits_{_{m=1}}^{M} {p_{mi}}\alpha_{gmn}^2}=\tfrac{\left({\sum\limits_{m=1}^{M}\sqrt{p_{mn}}\sum\limits_{g=1}^{G}x_{gn}\alpha_{gmn}}\right)^{2}}{\sigma^2+\sum\limits_{g=1}^{G}x_{gn}\sum\limits_{i=1}^{N}x_{gi}\sum\limits_{m=1}^{M}p_{mi}\beta_{mn}\alpha_{gmi}},~~~    ,1 \leq n\leq N.
\end{aligned}
\end{equation}
 \end{figure*}}

Instantaneous downlink channel state information (CSI) is needed at APs for beamforming in following downlink payload data transmission. So in the uplink training phase of this system, each user need to send pilot sequences simultaneously to APs for channel estimation on the same assigned time-slot. In this paper, the length of pilot sequence in group $g$ is $\tau_g$, which is the main overhead of downlink channel estimation at the users and increases linearly with the number of users in each group. To ensure the channel estimation accuracy, the length of pilot sequence is usually assumed no less than the number of users\cite{Ngo2017a, Li2018}. Note that user grouping can effectively increase the number of symbols for data transmission. We assume that the length of coherence interval is $\tau_c$ and the length of pilot sequence in group $g$ is $\tau_g$.

{The effect of coherent time on the performance of the proposed system mainly includes two aspects. One is the number of symbols for data transmission in each time-slot, and the other one is the data rate of users. In detail, considering the case that the number of users sharing the same time-slot is settled, the length of pilot sequence in this time-slot will be settled. Then the longer the coherence time, the longer is the efficient time of each channel estimation result and the length of coherence interval. The length of coherence interval is the sum of the length of pilot sequence and the number of symbols for data transmission in each time-slot. So the number of symbols for data transmission in each time-slot increases with the coherence time. On the other hand, the longer coherent time, the longer is the proportion of the symbols for data transmission. In each group or time-slot, the length of pilot sequences required for channel estimation increases linearly with the number of users sharing the same time-frequency resource. Without user grouping, the number of symbols for data transmission within a coherent interval will be only $\tau_c-\sum_{g=1}^{G}\tau_g$. After user grouping, the number of users in each time slot will decrease, and hence the length of pilot sequences required for channel estimation in each coherence interval will also decrease. The number of symbols for data transmission within a coherent interval is $\tau_c-\tau_g$. In other words, the length of pilot sequence in each time-slot will be reduced by user grouping. For example: In a downlink TDD based cell-free massive MIMO system with $\tau_c=100$ and 50 users, the length of pilot sequences required for channel estimation in each coherence interval is 50, and the proportion of the symbols for data transmission in each coherence interval is $50/100=0.5$. If we assign these users into $5$ groups, the users sharing the same coherence interval will be $10$, then the length of pilot sequences required for channel estimation in each coherence interval is 10, and the proportion of the symbols for data transmission in each coherence interval will up to $(100-10)/100=0.9$, which means that more time slots is used for downlink payload data transmission. Furthermore, the fewer users sharing the same coherence interval, the less interference that users will suffer and the lower decoding difficulty for the receiver.}

{We assume that the user grouping information (containing which group to access and the number of users in this group) has been acquired at each user.} Let $\bm{\psi}_{gn}\in \mathbb{C}^{\tau_g}$ denote the pilot sequence of the user $n$ if user $n$ is assigned into group $g$ and satisfies $\|\bm{\psi}_{gn}\|^2=1$. Then with a given user grouping matrix $\bm{x}=[x_{gn}]_{1\leq g \leq G, 1\leq n \leq N}$, where $x_{gn}$ is a 0-1 variable ($x_{gn}=1$ denotes that user $n$ is assigned into group $g$ and $x_{gn}=0$ denotes that user $n$ is not assigned into group $g$), the pilot signal that AP $m$ receives on time-slot $g$ is $\bm{y}^p_{mg}=$ $\sum_{n=1}^{N}x_{gn}\bm{\psi}_{gn}\sqrt{\rho_r\tau_g}h_{gmn}+{\mathfrak{n}^g_{m}}$, where $\rho_r$ is the power of pilot signal from users, and ${\mathfrak{n}^g_{m}}\sim (0, \sigma^2)$ is the Additive White Gaussian Noise (AWGN) received at APs. Let $\hat{h}_{gmn}$ denote the minimum mean square error (MMSE) estimate of $h_{gmn}$. Assuming that user $n$ is assigned into group $g$, we have {$\hat{h}_{gmn}$ $=\tfrac{\mathbb{E}[(\bm{\psi} _{gn}^{H}\bm{y}^p_{mg})^*h_{gmn}]}{\mathbb{E}[|\bm{\psi}_{gn}^{H}\bm{y}^p_{mg}|^2]}\bm{\psi}_{gn}^{H}\bm{y}^p_{mg}=\tfrac{\sqrt{\rho_r\tau_g}h_{gmn}}{\sigma^2+\rho_r\tau_g\beta_{gmn}}\bm{\psi}_{gn}^{H}\bm{y}^p_{mg},   1\leq m\leq M, 1\leq g\leq G, 1 \leq n\leq N.$} According to \cite{Nayebi2017}, its distribution is{\small
\begin{equation} \label{0000-1+1}
\setlength\abovedisplayskip{1pt}
\setlength\belowdisplayskip{1pt}
\hat{h}_{gmn}\sim \mathcal{C}\mathcal{N}(0, \alpha_{gmn}),~~~     1\leq m\leq M, 1\leq g\leq G, 1 \leq n\leq N.
\end{equation}}where $\alpha_{gmn}=\tfrac{\rho_r \tau_g \beta_{mn}^2}{\sigma^2+\rho_r \tau_g \beta_{mn}}$. In this paper, the length of pilot signal of users in group $g$, i.e., $\tau_g$, is equal to the number of users in group $g$. Then $\alpha_{gmn}$ can be rewritten as{\small
\begin{equation} \label{0000-1+1+1}
\setlength\abovedisplayskip{1pt}
\setlength\belowdisplayskip{1pt}
\alpha_{gmn}=\tfrac{\rho_r \sum _{i=1}^{N}x_{gi} \beta_{mn}^2}{\sigma^2+\rho_r \sum _{i=1}^{N}x_{gi} \beta_{mn}},~~     1\leq m\leq M, 1\leq g\leq G, 1 \leq n\leq N.
\end{equation}}Therefore, $\alpha_{gmn}$ is up to $\bm{x}$ and $\rho_r$ when large-scale fading is given. In addition, the distribution of channel estimation error is
{\small
\begin{equation} \label{0000-1+3}
\setlength\abovedisplayskip{1pt}
\setlength\belowdisplayskip{-10pt}
h_{gmn}-\hat{h}_{gmn}\sim \mathcal{C}\mathcal{N}(0, \beta_{mn}-\alpha_{gmn}),~~     1\leq m\leq M, 1\leq g\leq G, 1 \leq n\leq N.
\end{equation}}
\vspace{-0.2cm}
\subsection{Downlink Payload Data Transmission} \label{sec:02B}
After uplink training, APs will send the results of channel estimation to CPU as Fig.{\ref{fig:01}} shows. CPU calculates power control coefficients $p _{mn}$ (AP $m$ allocated to user $ n $) and sends these coefficients to each AP.
{ We w.l.o.g. assume that the users in group $g$ have been sorted and numbered by their channel gains. The index of user $n$ in its group is denoted by $\kappa_n$. Next we will introduce the conjugate beamforming (i.e., maximum-ratio transmission, MRT) and the ZF beamforming for precoding. First, we define the coefficients for beamforming as follows:{
\small
\begin{equation}\label{sec:02B00}
\setlength\abovedisplayskip{1pt}
\setlength\belowdisplayskip{1pt}
	%\begin{aligned}\nonumber
b_{gmn} =\!\!\left\{ \begin{array}{ll}
\!\!\!\!\hat{h}^*_{gmn}
\quad &\text{MRT}\\
\!\!\!\![\hat{\bm{H}}_g^*(\hat{\bm{H}}_g^T\hat{\bm{H}}_g^*)^{-1}]_{m\kappa_n}, ~\!\qquad\qquad\qquad &\text{ZF}
 \end{array}\right.,
\end{equation}}where $\bm{H}_g$ denotes the $M\times \sum _{i=1}^{N}x_{gi}$ channel coefficient matrix, $[\bm{H}_g]_{m\kappa_n}=h_{gmn}$. {It should be noted that, the real transmit power from AP $m$ to user $n$ is $p_{mn}|b_{gmn}|^2$, where $b_{gmn}$ is the beamforming coefficient and $p_{mn}$ is the power control coefficient. Although the value of $p_{mn}$ from different APs to each user are equal, the value of $b_{gmn}$ from different APs to each user are different. Therefore, the real transmit power from different APs to each user, i.e., $p_{mn}|b_{gmn}|^2$, are different.}
}

\subsubsection{Conjugate Beamforming} \label{sec:02B1}

The signal transmitted from each AP on time-slot $g$ after conjugate beamforming is given by ${t}_{gm}=\sum _{n=1}^{N}x_{gn}\sqrt {p_{mn}}{{b_{gmn}}}s_n$\cite{8599043,Zhang2019Onthe,8703420}, where $p _{mn}$ denotes power control coefficient of transmit power that AP $m$ allocated to user $ n $ , $s_n$ represents the transmitted symbol of user $n$. The signal received at user $n${, using MRT, is given by\footnote{{Although the strict phase-synchronization and -calibration between APs are not needed for non-coherent receivers, the coherent joint transmission can achieve higher spectral efficiency than non-coherent joint transmission. In the system model of this paper, the pilot sequences is sent by users and received by APs. So the channel estimation is only available at APs and the users only knows the large-scale channel gain. With coherent joint transmission, the user don't need to distinguish the signals from different APs. Based on the above considerations, the coherent joint transmission is adopted in this paper.}} {\small
\begin{equation}\label{0000+1}
\setlength\abovedisplayskip{1pt}
\setlength\belowdisplayskip{1pt}
	\begin{aligned}
	&{y_n^{MRT}}=\sum _{g=1}^{G}\sum_{m=1}^{M}x_{gn}h_{gmn}{t}_{gm}+\mathfrak{n}^g_{n}\\
&=\sum_{g=1}^{G}\sum_{m=1}^{M}x_{gn}h_{gmn}\sum _{i=1}^{N}x_{gi}\sqrt {p_{mi}}{b_{gmi}}s_i+\mathfrak{n}^g_{n},~~~     1 \leq n\leq N.
	\end{aligned}
\end{equation}}}where ${\mathfrak{n}^g_{m}}\sim (0, \sigma^2)$ is the AWGN received at users. For convenience, according to \cite{Nayebi2017}, we divide $y_n$ into five parts, including the desired signal of user $n$ $Y_{us}$, {interference from the desired signals of the other users} in the same group $Y_{in}$, the channel estimation error $Y_{ce}$, the lack of channel knowledge at user $Y_{lc}$ and noise $\mathfrak{n}^g_{n}$, and define the first four parts as follows. {\small
\begin{equation}\label{0000+6-1}\nonumber
\setlength\abovedisplayskip{1pt}
\setlength\belowdisplayskip{1pt}
\begin{aligned}
Y_{ce}=&\sum_{g=1}^{G}\sum_{m=1}^{M}\sum _{i=1}^{N}x_{gn}x_{gi}\sqrt {p_{mi}}(h_{gmn}-\hat{h}_{gmn}){\hat{h}_{gmi}^*}s_i,\\
Y_{us}=&\sum_{g=1}^{G}\sum_{m=1}^{M}x_{gn}\sqrt {p_{mn}}\mathbb{E}[|\hat{h}_{gmn}|^2]s_n,\\
	\end{aligned}
\end{equation}}

{\small
\begin{equation}\label{0000+6-1}\nonumber
\setlength\abovedisplayskip{1pt}
\setlength\belowdisplayskip{1pt}
\begin{aligned}
Y_{lc}=&\sum_{g=1}^{G}\sum_{m=1}^{M}x_{gn}\sqrt {p_{mn}}(|\hat{h}_{gmn}|^2-\mathbb{E}[|\hat{h}_{gmn}|^2])s_n,\\
Y_{in}=&sum_{g=1}^{G}\sum_{m=1}^{M}\sum _{^{i=1}_{i\neq n}}^{N}x_{gn}x_{gi}\sqrt {p_{mi}}\hat{h}_{gmn}{\hat{h}_{gmi}^*}s_i.
	\end{aligned}
\end{equation}}Then (\ref{0000+1}) can be rewritten as ${y_n^{MRT}}=Y_{us}+Y_{in}+Y_{ce}+Y_{lc}+\mathfrak{n}^g_{n}$. As these five parts are mutually uncorrelated, the lower bound of SINR achievable to user $n$ is \cite{Nayebi2017}, {where the time interval is the coherence time (a time slot), hence the noise in this expression is one $\sigma^2$. It should be noted that $\sum\limits_{g=1}^{G}x_{gn}=1$, for $\forall n$. Hence, the SINR of users in different groups will not affect with each other.}

Obviously, {interference from the desired signals of the other users} in the same group will be reduced as the number of users in each group decreases, and the complexity of decoding at the receiver will be relatively reduced. {To satisfy the user QoS requirements, the achievable SINR of each user should be constrained by
$\mathrm{SINR}_n^{MRT} \geq \gamma_n ,       ~~~ 1 \leq n\leq N,1 \leq g\leq G$  where $\gamma_n=2^{GR_n^{target}\frac{\tau_c}{\tau_c-\tau_g}}-1$ is the minimal SINR in the transmit time slot of user $n$ to achieve its target data rate $R_n^{target}$, and $\frac{\tau_c}{\tau_c-\tau_g}$ is the ratio of the length of data in each coherence interval after grouping and that of data in each coherence interval without grouping.}

{\subsubsection{Zero-Forcing Precoder} \label{sec:02B2}

With zero-forcing precoder, the signal transmitted from each AP on time-slot $g$ is given by ${t}_{gm}=\sum _{n=1}^{N}x_{gn}\sqrt {p_{n}} b_{gmn} s_n$, where $p_{n}$ is the power allocation coefficient of user $n$ under the assumption of $p_{mn}=p_{n}$, for $\forall m$. The signal received at user $n$, using ZF, is given by $y^{ZF}_n=\sum _{g=1}^{G}\sum_{m=1}^{M}x_{gn}h_{gmn}{t}_{gm}+\mathfrak{n}^g_{n},~~~     1 \leq n\leq N.$

Since interference from the desired signals of the other users in the same group has been eliminated by zero-forcing precoder, the lower bound of SINR achievable to user $n$ is \cite{Femenias2020,Nayebi2017,Nayebi2016} $\mathrm{SINR}_{n}^{ZF}=\tfrac{p_{n}}{\sigma^2+\sum _{g=1}^{G}x_{gn}\sum_{i=1}^{N}x_{gi}p_{i}\eta_{ni}}$,
where $\eta_{ni}$ is the $\kappa_i$-th element of the following vector:$\bm{\eta}_{n}=diag\{\mathbb{E}((\hat{\bm{H}}_g^T\hat{\bm{H}}_g^*)^{-1}\hat{\bm{H}}_g^T\mathbb{E}(\hat{\bm{h}}_{g\kappa_n}^*\hat{\bm{h}}_{g\kappa_n}^T)\hat{\bm{H}}_g^*(\hat{\bm{H}}_g^T\hat{\bm{H}}_g^*)^{-1})\}$, where $\hat{\bm{h}}_{g\kappa_n}^T=[\hat{{h}}_{g1n},\ldots\hat{{h}}_{gmn}\ldots\hat{{h}}_{gMn}]$, and $\mathbb{E}(\hat{\bm{h}}_{g\kappa_n}^*\hat{\bm{h}}_{g\kappa_n}^T)$ is a diagonal matrix with $(\beta_{mn}-\alpha_{gmn})$ on its m-th diagonal element, which has been proved in \cite{Nayebi2016}. The value of $\eta_{ni}$ can be obtained using exponential smoothing as stated in \cite{Nayebi2017}. { Exponential smoothing is a method to predict the future value of a variable by weighting its past values considering the change trend of its value. In this paper, the historical values of $\eta_{ni}$ can be obtained from previous channel estimation results, hence we can obtain the predicted value of $\eta_{ni}$ by weighting its historical values. We assume that the accurate and the estimated value of the current $\eta_{ni}$ is $\eta_{ni}^t$ and $\eta_{ni}^{(t)}$, respectively. and $\eta_{ni}^{t-1}$ and $\eta_{ni}^{(t-1)}$ denote the accurate and the estimated value of the $(t-1)$-th $\eta_{ni}$. Then $\eta_{ni}^{(t)}=w\eta_{ni}^{t-1}+(1-w)\eta_{ni}^{(t-1)}$, where $w$ is a constant between 0 and 1.}

To get a similar form of $\mathrm{SINR}_{n}^{MRT}$ in (\ref{0001}), $\mathrm{SINR}_{n}^{ZF}$ can be rewritten as:{\small
\setlength\abovedisplayskip{1pt}
\setlength\belowdisplayskip{1pt}
\begin{equation}\label{sec:02B203+}
\begin{aligned} &\mathrm{SINR}_{n}^{ZF}=\tfrac{\left({\sum\limits_{m=1}^{M}\tfrac{1}{M}\sqrt{p_{mn}}}\right)^{2}}{\sigma^2+\sum\limits_{g=1}^{G}x_{gn}\sum\limits_{i=1}^{N}x_{gi}\sum\limits_{m=1}^{M}\tfrac{1}{M}p_{mi}\eta_{ni}},~1 \leq n\leq N\Bigg|p_{mn}=p_{n}.
\end{aligned}
\end{equation}}
}
\vspace{-0.5cm}
\subsection{Problem Formulation}
{According to (\ref{0001}) and (\ref{sec:02B203+}), the values of $\mathrm{SINR}_n^{MRT}$ and $\mathrm{SINR}_n^{ZF}$ are} up to the user grouping matrix $\bm{x}=[x_{gn}]_{1\leq g \leq G, 1\leq n \leq N,}$, power allocation matrix $\bm{p}=[p_{mn}]_{1\leq m \leq M, 1\leq n \leq N}$, and large-scale fading matrix $\bm{\beta}=[\beta_{mn}]_{1\leq m \leq M, 1\leq n \leq N}$ (The value of $\alpha_{gmn}$ is up to $\bm{x}$ if $\rho_{\tau}$ and $\bm{\beta}$ are given as stated in subsection \ref{sec:02A}). We try to optimize the user grouping and power allocation strategy to minimize the total transmit power with known large-scale fading under QoS constraints, i.e., target SINR. { We formulate this joint power allocation and user grouping problem as $\mathcal{P}1$.
{\small
\setlength\abovedisplayskip{1pt}
\setlength\belowdisplayskip{1pt}
\begin{subequations}\label{0003}
	\begin{align}
    \mathcal{P}1:~	\min_{\bm{x},\bm{p}} ~~ &P_t={\sum\limits_{m=1}^{M}\sum\limits_{n=1}^{N}p_{mn}\sum\limits_{g=1}^{G}x_{gn}\varphi_{gmn}}\label{0003a}\\
    \text{s.t.}~~~~&\tfrac{\left({\sum\limits_{m=1}^{M}\sqrt{p_{mn}}\sum\limits_{g=1}^{G}x_{gn}\vartheta_{gmn}}\right)^{2}}{\sigma^2+\sum\limits_{g=1}^{G}x_{gn}\sum\limits_{i=1}^{N}x_{gi}\sum\limits_{m=1}^{M}p_{mi}\upsilon_{gmni}}\geq \gamma_n,~~~     1 \leq n\leq N,\label{0003b}\\
    &p_{mn}\geq 0,~~~~1 \leq n\leq N, 1 \leq m\leq M,\label{0003c}\\
    &\sum\limits_{g=1}^{G}x_{gn}=1,~~~ x_{gn}\in \{0,1\},1 \leq n\leq N,\label{0003d}\\
    &p_{mn}=p_{m'n},~~~1\leq m,M'\leq M\Bigg|ZF\label{0003e}\\
    &\text{(\ref{0000-1+1+1}),(\ref{sec:02B04+})-(\ref{sec:02B04+++})}\nonumber.
	\end{align}
\end{subequations}}}where (\ref{0003d}) means that each user should be assigned into only one group, { (\ref{0003e}) exists when ZF beamforming is chosen. Like (\ref{sec:02B00}), some variables are defined as follows:{\small
\setlength\abovedisplayskip{1pt}
\setlength\belowdisplayskip{1pt}
\begin{equation}\label{sec:02B04+}
\varphi_{gmn} =\left\{ \begin{array}{ll}
\alpha_{gmn}
\quad &\text{MRT}\\
\text{$[diag\{\mathbb{E}((\hat{\bm{H}}_g^T\hat{\bm{H}}_g^*)^{-1}\hat{\bm{h}}_{[gm]}^*\hat{\bm{h}}_{[gm]}^T(\hat{\bm{H}}_{g}^T\hat{\bm{H}}_{g}^*)^{-1})\}]_{\kappa_n}$} &\text{ZF}
 \end{array}\right.
\end{equation}},where $\hat{\bm{h}}_{gm}$ is the m-th row of $\hat{\bm{H}}_g$.
{\small
\setlength\abovedisplayskip{0pt}
\setlength\belowdisplayskip{0pt}
\begin{equation}\label{sec:02B04++}
\vartheta_{gmn} =\left\{ \begin{array}{ll}
\alpha_{gmn}
\quad &\text{MRT}\\
\tfrac{1}{\sqrt{M}}, ~\qquad\qquad\qquad &\text{ZF}
 \end{array}\right.,
\end{equation}}{\small
\setlength\abovedisplayskip{0pt}
\setlength\belowdisplayskip{-10pt}
\begin{equation}\label{sec:02B04+++}
\upsilon_{gmni} =\left\{ \begin{array}{ll}
\beta_{mn}\alpha_{gmi}
\quad &\text{MRT}\\
\tfrac{1}{\sqrt{M}}\eta_{ni}, ~\qquad\qquad\qquad &\text{ZF}
 \end{array}\right.,
\end{equation}}}\quad To solve this MINLP problem, we first define a $M\times N$ matrix $\bm{q}$$=[q_{mn}]_{1 \leq m\leq M, 1 \leq n\leq N}$, where $q_{mn}=\sqrt{p_{mn}}$, and convert problem $\mathcal{P}1$ into problem $\mathcal{P}2$:
{
{\small
\setlength\abovedisplayskip{3pt}
\setlength\belowdisplayskip{3pt}
\begin{subequations}\label{0003+1}
	\begin{align}
	\mathcal{P}2:~\min_{\bm{x},\bm{q}} ~~ &P_t={\sum\limits_{m=1}^{M}\sum\limits_{n=1}^{N}q_{mn}^2\sum\limits_{g=1}^{G}x_{gn}\varphi_{gmn}},\label{0003+1a}\\
    \text{s.t.}~~~&\left({\sigma^2+\sum\limits_{g=1}^{G}x_{gn}\sum\limits_{i=1}^{N}x_{gi}\sum\limits_{m=1}^{M}q_{mi}^2\upsilon_{gmni}} \right)^{\tfrac{1}{2}} \gamma_n^{\tfrac{1}{2}}\nonumber\\-
    & {{\sum\limits_{m=1}^{M}q_{mn}\sum\limits_{g=1}^{G}x_{gn}\vartheta_{gmn}}}\leq 0,~~~     1 \leq n\leq N,\label{0003+1b}\\
    &q_{mn}\geq 0,~~~~1 \leq n\leq N, 1 \leq m\leq M,\label{0003+1c}\\
    &q_{mn}=q_{m'n},~~~1\leq m,M'\leq M\Bigg|ZF\label{0003+1d}\\
    &\text{(\ref{0000-1+1+1}),(\ref{sec:02B04+})-(\ref{sec:02B04+++}), (\ref{0003d})}.\nonumber
	\end{align}
\end{subequations}}}A key motivation of this conversion is to convert constraints (\ref{0003b}) into convex constraints (\ref{0003+1b}) with given user grouping matrix $\bm{x}$, which will be introduced in the next section. Problem $\mathcal{P}2$ is still hard to solve, we try to solve it with an iterative method based on GBD method \cite{8789693,8374812}.

\section{Problem Analysis and Solutions}

We first decompose this problem into a primal problem: power allocation problem and a master problem: user grouping problem according to GBD method \cite{8789693} \cite{8374812}. Then according to the basic principle of GBD method, the MINLP problem can be solved by solving these two problems iteratively \cite{Xiang2017}. In each iteration, the upper bound and the lower bound of the problem can be updated, and the gap among the upper and lower bound is shrunk \cite{8384307}.

\subsection {Primal Problem: Power allocation problem}

Power allocation problem $\mathcal{S}^{(k)}$ is given by fixing the user grouping matrix to $\bm{x}^{(k)}$:{{\small
\begin{subequations}\label{0005}
	\begin{align}
	\mathcal{S}^{(k)}:~\min_{\bm{q}} ~ &P_t={\sum\limits_{m=1}^{M}\sum\limits_{n=1}^{N}q_{mn}    ^2 \sum\limits_{g=1}^{G}x^{(k)}_{gn}\varphi^{(k)}_{gmn}}\label{0005a}\\
    \text{s.t.}~~&\left({\sigma^2+\sum\limits_{g=1}^{G}x^{(k)}_{gn}\sum\limits_{i=1}^{N}x^{(k)}_{gi}\sum\limits_{m=1}^{M}q_{mi}^2\upsilon_{gmni}^{(k)}} \right)^{\tfrac{1}{2}} \gamma_n^{\tfrac{1}{2}}\nonumber\\-
    &{{\sum\limits_{m=1}^{M}q_{mn}\sum\limits_{g=1}^{G}x^{(k)}_{gn}\vartheta_{gmn}^{(k)}}}\leq 0,~~    1 \leq n\leq N,\label{0005b}\\
    &\text{(\ref{0000-1+1+1}),(\ref{0003+1c}),(\ref{sec:02B04+})-(\ref{sec:02B04+++}),(\ref{0003+1d})}.\nonumber
	\end{align}
\end{subequations}}}Since the objective function (\ref{0005a}) is convex, the SINR constraints (\ref{0005b}) are second order cone (SOC) constraints, the constraints (\ref{0003+1c}) and (\ref{0003+1c}) are linear and the other constraints are not related to the value of $\bm{q}$, problem $\mathcal{S}^{(k)}$ is a convex problem.

Since problem $\mathcal{S}^{(k)}$ is a convex problem, we can solve it by the interior point method. In addition, problem $\mathcal{S}^{(k)}$ is given by fixing the user grouping matrix to $\bm{x}^{(k)}$. Hence, there are two cases about this problem, feasible and infeasible. These cases of problem $\mathcal{S}^{(k)}$ are discussed as follows:

\textbf{Feasible Case:} We first define the partial Lagrangian function of problem $\mathcal{S}^{(k)}$ \cite{8123914}:{{\small
		\begin{equation}\label{0016}
\begin{aligned}
&\mathcal{L}(\bm{q},\bm{\lambda}, \bm{x}^{(k)})={\sum\limits_{m=1}^{M}\sum\limits_{n=1}^{N}q_{mn}^2\sum\limits_{g=1}^{G}x^{(k)}_{gn}\varphi^{(k)}_{gmn}}\\
&+\sum\limits_{n=1}^{N}\lambda_n\left(\left({\sigma^2+\sum\limits_{g=1}^{G}x^{(k)}_{gn}\sum\limits_{i=1}^{N}x^{(k)}_{gi}\sum\limits_{m=1}^{M}q_{mi}^2\upsilon^{(k)}_{gmni}} \right)^{\tfrac{1}{2}} \gamma_n^{\tfrac{1}{2}}\right.\\ &-\left. {{\sum\limits_{m=1}^{M}q_{mn}\sum\limits_{g=1}^{G}x^{(k)}_{gn}\vartheta^{(k)}_{gmn}}}\right),
\end{aligned}
\end{equation}}}where the Lagrangian multipliers $\bm{\lambda}=[\lambda_{mn}]_{1 \leq m\leq M, 1 \leq n\leq N}$ correspond to constraints (\ref{0005b}) and satisfy $\lambda_{mn}\geq 0$, $1 \leq m\leq M,~1 \leq n\leq N$. The dual problem of problem $\mathcal{S}^{(k)}$ can be obtained as stated in Lemma \ref{lem1}.

\begin{lem1}\label{lem1}

Problem $\mathcal{S}^{(k)}$ is equivalent to its dual problem $\mathcal{D}_S^{(k)}$ as follows \cite{8543658}.
{\small
\setlength\abovedisplayskip{3pt}
\setlength\belowdisplayskip{3pt}
\begin{subequations}\label{0017}
	\begin{align}
	\mathcal{D}_S^{(k)}:~\max_{\bm{\lambda}} ~~ &\inf_{\bm{q}}~\mathcal{L}(\bm{q},\bm{\lambda}, \bm{x}^{(k)})\label{0017a}\\
    \text{s.t.}~~~~&\lambda_{mn}\geq 0,~~~~1 \leq m\leq M,~1 \leq n\leq N,\label{0017b}\\
    &\text{{(\ref{0000-1+1+1}),(\ref{0003+1c}),(\ref{0003+1d}).}}\nonumber\nonumber
	\end{align}
\end{subequations}}\end{lem1}

\begin{proof}

It is obvious that there exists a strictly feasible point for convex problem $\mathcal{S}^{(k)}$ (feasible case), thus Slater's condition is satisfied \cite{7995020} \cite{6355622}. Hence, strong duality holds for problem $\mathcal{S}^{(k)}$ and its dual problem \cite{8012535}.\end{proof}

\textbf{Infeasible Case:}
According to Lemma \ref{lem1}, we can get an upper bound of problem $\mathcal{P}2$. Then if problem $\mathcal{S}^{(k)}$ is infeasible, which means that constraint (\ref{0005b}) can not be satisfied no matter how we allocate power, we try to find the power allocation strategy that is close to constraint (\ref{0005b}). By relaxing constraint (\ref{0005b}) with a violation variable $\phi$, we can get the following problem $\mathcal{S}2^{(k)}$ as well as its dual problem $\mathcal{D}_{S2}^{(k)}$ \cite{geoffrion1972generalized}. In problem $\mathcal{S}2^{(k)}$, we try to minimize the gap between the left and right sides of (\ref{0005b}).{\small
{\setlength\abovedisplayskip{3pt}
\setlength\belowdisplayskip{3pt}
\begin{subequations}\label{0018}
	\begin{align}
	\mathcal{S}2^{(k)}:~\min_{\bm{q},\phi} ~ &\phi\label{0018a}\\
    \text{s.t.}~~&\left({\sigma^2+\sum\limits_{g=1}^{G}x^{(k)}_{gn}\sum\limits_{i=1}^{N}x^{(k)}_{gi}\sum\limits_{m=1}^{M}q_{mi}^2\upsilon^{(k)}_{gmni}} \right)^{\tfrac{1}{2}} \gamma_n^{\tfrac{1}{2}}\\-& {{\sum\limits_{m=1}^{M}q_{mn}\sum\limits_{g=1}^{G}x^{(k)}_{gn}\vartheta^{(k)}_{gmn}}}\leq \phi,~~1 \leq n\leq N,\label{0018b}\\
    &\phi\geq 0,\label{0018c}\\
    &\text{(\ref{0000-1+1+1}),(\ref{0003+1c}),(\ref{0003+1d}).}\nonumber
	\end{align}
\end{subequations}}}We define the partial Lagrangian function of $\mathcal{S}2^{(k)}$ as follows:{\small
{
		\begin{equation}\label{0019}
\begin{aligned}
\mathcal{L}'(\bm{q}, \bm{\nu}, \bm{x}^{(k)})&=\sum\limits_{n=1}^{N}\nu_n\left(\left({\sigma^2+\sum\limits_{g=1}^{G}x^{(k)}_{gn}\sum\limits_{i=1}^{N}x^{(k)}_{gi}\sum\limits_{m=1}^{M}q_{mi}^2\upsilon^{(k)}_{gmni}} \right)^{\tfrac{1}{2}} \gamma_n^{\tfrac{1}{2}}\right.\\-&\left. {{\sum\limits_{m=1}^{M}q_{mn}\sum\limits_{g=1}^{G}x^{(k)}_{gn}\vartheta^{(k)}_{gmn}}}\right),
\end{aligned}
		\end{equation}}}where the Lagrangian multipliers $\bm{\nu}$ correspond to constraints $(\ref{0018b})$ and satisfy $\nu_{mn}\geq 0$, $1 \leq m\leq M,~1 \leq n\leq N$. Like problem $\mathcal{S}^{(k)}$, we can also get the optimal solutions $\bm{q}^{(k)}$ and the dual solutions $\bm{\nu}^{(k)}$ by the interior point method. In addition, Lemma \ref{lem1+} is obtained.

\vspace{-0.2cm}
\begin{lem1}\label{lem1+}

Problem $\mathcal{S}2^{(k)}$ is equivalent to its dual problem $\mathcal{D}_{S2}^{(k)}$ as follows.{\small
\setlength\abovedisplayskip{1pt}
\setlength\belowdisplayskip{0pt}
\begin{subequations}\label{0019+1}
	\begin{align}
	\mathcal{D}_{S2}^{(k)}:~\max_{\bm{\nu}} ~~ &\inf_{\bm{q}, \phi}\mathcal{L}'(\bm{q}, \bm{\nu}, \bm{x}^{(k)})+\phi-\sum\limits_{n=1}^{N}\nu_n\phi\label{0019+1a}\\
    \text{s.t.}~~~~&\nu_{mn}\geq 0,~~~~1 \leq m\leq M,~1 \leq n\leq N,\label{0019+1b}\\
    &\text{(\ref{0018c}), ~{(\ref{0000-1+1+1}), ~(\ref{0003+1c}),(\ref{0003+1d}).}}\nonumber
	\end{align}
\end{subequations}}\end{lem1}\begin{proof}
Since the objective function (\ref{0019+1a}) is convex and all the constraints are linear, problem $\mathcal{S}2^{(k)}$ is convex. In addition, for any $\bm{q}\succ 0$ and any ${\phi}$ satisfies (\ref{0018b}), $\{\bm{q},\phi\}$ is feasible for convex problem $\mathcal{S}2^{(k)}$, thus Slater's condition is satisfied. Hence, strong duality holds for problem $\mathcal{S}2^{(k)}$ and its dual problem.\end{proof}

\vspace{-0.4cm}
\subsection{Master problem: User grouping problem}We write the master user grouping problem $\mathcal{M}1$ as follows:
{\setlength\abovedisplayskip{1pt}
\setlength\belowdisplayskip{1pt}
\begin{subequations}\label{0019+2}\small
	\begin{align}\nonumber
	\mathcal{M}1:~\min_{\bm{x}} ~~ &\text{{f}}(\bm{x})\\
    \text{s.t.}~~~~ &\bm{x}\in \{\bm{x} \text{$^{(k)}$}|\text{$\mathcal{S}^{(k)}$ is feasible}\},~~~     1 \leq n\leq N,\label{0019+2a}\\
    &\text{(\ref{0003d})}.\nonumber
	\end{align}
\end{subequations}}where function {f}($\bm{x}$) returns the optimal value of problem $\mathcal{S}^{(k)}|_{\bm{x}^{(k)}=\bm{x}}$.

In problem $\mathcal{M}1$, (\ref{0019+2a}) is not in the explicit form. Therefore, to apply GBD method, we convert the master problem $\mathcal{M}1$ into an explicit form in the following lemma.

\begin{lem1}\label{lem2}

Problem $\mathcal{M}1$ is equivalent to the following problem $\mathcal{M}2$.{\small
\setlength\abovedisplayskip{1pt}
\setlength\belowdisplayskip{1pt}
\begin{subequations}\label{0020}
	\begin{align}\nonumber
	\mathcal{M}2:~\min_{\bm{x}} ~~ &\xi\\
    \text{s.t.}~~~~&\min_{\bm{q}\succeq 0}\mathcal{L}(\bm{q},\bm{\lambda}, \bm{x})\leq \xi, \forall \bm{\lambda} \succeq 0,\label{0020a}\\
    &\min_{\bm{q}\succeq 0}\mathcal{L}'(\bm{q}, \bm{\nu}, \bm{x}) \leq 0,\forall \bm{\nu} \succeq 0: \sum_{n=1}^{N}\nu_{n}=1,\label{0020b}\\
    &\text{(\ref{0003d}))}.\nonumber
	\end{align}
\end{subequations}}\end{lem1}
\begin{proof}
Since the constraints (\ref{0005b}) are convex, according to Theorem 2.2 in \cite{Angeles2012}, the constraints (\ref{0019+2a}) are equivalent to (\ref{0020b}). Then according to Lemma \ref{lem1}, we have:
{\small
\setlength\abovedisplayskip{1pt}
\setlength\belowdisplayskip{1pt}
		\begin{equation}\label{0021}
\begin{aligned}
\text{{f}}(\bm{x})=\max_{\bm{\lambda\succeq 0}}~\min_{\bm{q}\succeq 0}\mathcal{L}(\bm{q},\bm{\lambda}, \bm{x}).
\end{aligned}
		\end{equation}} Therefore, the following two problems are equivalent.

{\small\setlength\abovedisplayskip{1pt}
\setlength\belowdisplayskip{1pt}
\begin{subequations}\label{0021+1}
	\begin{align}\nonumber
	\min_{\bm{x}} ~~ &\text{{f}}(\bm{x})\\
    \text{s.t.}~~~~&\text{(\ref{0003d})}.\nonumber
	\end{align}
\end{subequations}}
and{\small\setlength\abovedisplayskip{1pt}
\setlength\belowdisplayskip{1pt}
\begin{subequations}\label{0021+2}
	\begin{align}\nonumber
	\min_{\bm{x}} ~~ &\xi\\\nonumber
\text{s.t.}~~~~&\min_{\bm{q}\succeq 0}\mathcal{L}(\bm{q},\bm{\lambda}, \bm{x})\leq \xi, \forall \bm{\lambda} \succeq 0,\\\nonumber
    &\text{(\ref{0003d})}.
	\end{align}
\end{subequations}}Thus, the proof of Lemma \ref{lem2} is concluded.\end{proof}

The constraints (\ref{0020a}) and (\ref{0020b}) are composed of an infinite number of constraints ($\bm{q}$ is a matrix composed of continuous variables), which makes problem $\mathcal{M}2$ hard to solve. Next we settle the Lagrangian multipliers $\bm{\lambda}$ and $\bm{\nu}$ to make problem $\mathcal{M}2$ more explicit by Lemma \ref{lem3}.

\begin{lem1}\label{lem3}

If $\mathcal{S}^{(k)}$ is feasible, and both its optimal solution $\bm{q}^{(k)}$ and the dual solution $\bm{\lambda}^{(k)}$ have been obtained, we have:{\small
		\begin{equation}\label{0022}
\begin{aligned}
\min_{\bm{q}\succeq 0}\mathcal{L}(\bm{q},\bm{\lambda}^{(k)}, \bm{x})=\mathcal{L}(\bm{q}^{(k)},\bm{\lambda}^{(k)}, \bm{x}).\\
\end{aligned}
		\end{equation}}If $\mathcal{S}^{(k)}$ is infeasible, and the optimal solution $\bm{q}^{(k)}$ as well as the dual solution $\bm{\nu}^{(k)}$ of problem $\mathcal{S}2^{(k)}$ have been obtained, the following equations are equivalent.
{\small
\setlength\abovedisplayskip{1pt}
\setlength\belowdisplayskip{1pt}
		\begin{equation}\label{0023}
\begin{aligned}
\min_{\bm{q}\succeq 0}\mathcal{L}'(\bm{q}, \bm{\nu}^{(k)}, \bm{x})\leq 0: \sum_{n=1}^{N}\nu_{n}^{(k)}=1,\\
\end{aligned}
		\end{equation}}and
{\small
\setlength\abovedisplayskip{1pt}
\setlength\belowdisplayskip{1pt}
		\begin{equation}\label{0024}
\begin{aligned}
\mathcal{L}'(\bm{q}^{(k)}, \bm{\nu}^{(k)}, \bm{x})\leq 0.\\
\end{aligned}
		\end{equation}}\end{lem1}
\begin{proof}

If $\mathcal{S}^{(k)}$ is feasible, according to (\ref{0017}), (\ref{0022}) is tenable.

If $\mathcal{S}^{(k)}$ is infeasible, according to (\ref{0019}), we have:
{\small
\setlength\abovedisplayskip{1pt}
\setlength\belowdisplayskip{1pt}
		\begin{equation}\label{0025}
\begin{aligned}
&\tfrac{\partial(\mathcal{L}'(\bm{q}, \bm{\nu}^{(k)}, \bm{x})+\phi-\sum\limits_{n=1}^{N}\nu_n^{(k)}\phi}{\partial \phi}=1-\sum\limits_{n=1}^{N}\nu_n^{(k)}=0\\
\end{aligned}
\end{equation}}Then according to (\ref{0018}) and (\ref{0019+1}) we have{\small
\setlength\abovedisplayskip{1pt}
\setlength\belowdisplayskip{1pt}
		\begin{equation}\label{0026}
\begin{aligned}
(\bm{q}^{(k)},\phi)=&{arg\min_{\bm{q}\succeq 0, \phi\geq 0}}~\mathcal{L}'(\bm{q}, \bm{\nu}^{(k)}, \bm{x})+\phi-\sum\limits_{n=1}^{N}\nu_n^{(k)}\phi\\
=&{arg\min_{\bm{q}\succeq 0, \phi\geq 0}}~\mathcal{L}'(\bm{q}, \bm{\nu}^{(k)}, \bm{x})+\phi(1-\sum\limits_{n=1}^{N}\nu_n^{(k)}),\\
\end{aligned}
		\end{equation}}Combining (\ref{0026}) with (\ref{0025}), we have
{\small
\setlength\abovedisplayskip{1pt}
\setlength\belowdisplayskip{1pt}
		\begin{equation}\label{0027}
\begin{aligned}
\bm{q}^{(k)}=&{arg\min_{\bm{q}\succeq 0}~\mathcal{L}'(\bm{q}, \bm{\nu}^{(k)}, \bm{x})}.
\end{aligned}
		\end{equation}}where $\sum_{n=1}^{N}\nu_{n}^{(k)}=1$. Hence, (\ref{0023}) and (\ref{0024}) are equivalent.
The proof of Lemma \ref{lem3} is concluded.\end{proof}

According to Lemma \ref{lem3}, we can relax problem $\mathcal{M}2$ into problem $\mathcal{M}3$ by calculating the optimal solution $\bm{q}^{(k)}$ and the dual solution $\bm{\lambda}^{(k)}$ of feasible problem $\mathcal{S}^{(k)}$, and we can relax problem $\mathcal{M}2$ into problem $\mathcal{M}3$ by calculating the optimal solution $\bm{q}^{(k)}$ and the dual solution $\bm{\nu}^{(k)}$ of problem $\mathcal{S}2^{(k)}$ when problem $\mathcal{S}^{(k)}$ is infeasible{\small
\setlength\abovedisplayskip{1pt}
\setlength\belowdisplayskip{1pt}
\begin{subequations}\label{0028}
	\begin{align}\nonumber
	\mathcal{M}3:~\min_{\bm{x}} ~~ &\xi\\
    \text{s.t.}~~~~&\mathcal{L}(\bm{q}^{(k_1)},\bm{\lambda}^{(k_1)}, \bm{x})\leq \xi,~~k_1\in \{k|\text{$\mathcal{S}^{(k)}$ is feasible}\}\label{0028b}\\
    &\mathcal{L}'(\bm{q}^{(k_2)}, \bm{\nu}^{(k_2)}, \bm{x})\leq 0,~~k_2\in \{k|\text{$\mathcal{S}^{(k)}$ is infeasible}\}\\
    &\text{(\ref{0003d})}.\nonumber
	\end{align}
\end{subequations}}Note that problem $\mathcal{M}3$ is more explicit than problem $\mathcal{M}1$.

\begin{lem1}\label{lem4}

The optimal value of problem $\mathcal{M}3$ is a lower bound of problem $\mathcal{P}2$.

\end{lem1}

\begin{proof}

The proof is stated in Remark 2.3 of \cite{Angeles2012} in detail.\end{proof}

\begin{algorithm}[htbp]\label{alg1}\small
\setlength\abovedisplayskip{1pt}
\setlength\belowdisplayskip{0pt}
 \caption{GBD Based Joint Power Allocation and User Grouping Algorithm (GPGA)}\label{alg:2}
  \KwIn{Variance of Channel Estimation $\bm{\alpha}$, Large-scale Fading $\bm{\beta}$, }

  \KwOut{Power Allocation Matrix $\bm{q}$, Grouping Matrix $\bm{x}$}
  \For{$n=1 \textrm{ to }N$}{
         $g=mod(n,G)+1$;         $x_{gn}^1=1$
        }
        $k=1$;Create problem $\mathcal{M}3$ without feasi-constraint or infeasi-constraint;

  \Repeat{$b_u-b_l\leq\delta$ or $b_u$ won't change or number of iterations exceeds N}{
        Solve power allocation problem $\mathcal{S}^{(k)}$ by the interior point method;

        \eIf{$\mathcal{S}^{(k)}$ is feasible.}{
          Calculate the optimal solution $\bm{q}^{(k)}$, optimal value $P_t^{(k)}={\sum\limits_{m=1}^{M}\sum\limits_{n=1}^{N}q_{mn}^2\sum\limits_{g=1}^{G}x^{(k)}_{gn}{\varphi_{gmn}^{(k)}}}$ and the dual solution $\bm{\lambda}^{(k)}$ of problem $\mathcal{S}^{(k)}$;

          Add constraint $\mathcal{L}(\bm{q}^{(k)}, \bm{\lambda}^{(k)}, \bm{x})\leq \xi$ to the relaxed master problem $\mathcal{M}3$;

          Update upper bound $b_u=P_t^{(k)}$;

        }
        {
          Calculate the optimal solution $\bm{q}^{(k)}$, $P_t^{(k)}={\sum\limits_{m=1}^{M}\sum\limits_{n=1}^{N}q_{mn}^2\sum\limits_{g=1}^{G}x^{(k)}_{gn}{\varphi_{gmn}^{(k)}}}$ and the dual solution $\bm{\nu}^{(k)}$ of problem $\mathcal{S}2^{(k)}$ by the interior point method;

          Add constraint $\mathcal{L}'(\bm{q}^{(k)}, \bm{\nu}^{(k)}, \bm{x})\leq 0$ to the relaxed master problem $\mathcal{M}3$;
        }

        Solve the relaxed master problem $\mathcal{M}3$ to get new grouping matrix $\bm{x}^{(k+1)}$ and its optimal value $\xi_{min}$;

        Update lower bound $b_l=\xi_{min}$;        $k=k+1$;

          }

\end{algorithm}

Since the upper bound and lower bound of problem $\mathcal{P}2$ can be obtained by solving problem $\mathcal{S}^{k}$ and problem $\mathcal{M}3$, respectively, we propose a joint power allocation and user grouping algorithm (GPGA) based on GBD as shown in Algorithm \ref{alg1}. In this algorithm, steps 1-4 is the initialization of grouping matrix $\bm{x}^{(k)}$. Then according to the feasibility of problem $\mathcal{S}^{(k)}$ and Lemma \ref{lem3}, we add constraint $\mathcal{L}(\bm{q}^{(k)}, \bm{\lambda}^{(k)}, \bm{x})\leq \xi$ or $\mathcal{L}'(\bm{q}^{(k)}, \bm{\nu}^{(k)}, \bm{x})\leq 0$, which we called them feasi-constraint and infeasi-constraint, respectively, to relaxed master problem $\mathcal{M}3$ and update the upper bound and lower bound of problem $\mathcal{P}2$. A new user grouping matrix $\bm{x}^{(k+1)}$ is obtained by solving problem $\mathcal{M}3$. Next, matrix $\bm{x}^{(k)}$ is updated by matrix $\bm{x}^{(k+1)}$. Steps 6-19 is repeated until the gap between the upper bound and lower bound of problem $\mathcal{P}2$ is less than $\delta$, i.e., $\delta$-optimal solution.

\begin{pro1}
\label{pro2}

Algorithm \ref{alg1} is bound to stop in finite steps for any given $\delta>0$.

\end{pro1}

\begin{proof}

After each iteration of 6-19 in Algorithm \ref{alg1}, the upper bound of problem $\mathcal{P}2$ is nonincreasing and the lower bound of problem $\mathcal{P}2$ is nondecreasing. Therefore, the gap between the upper bound and lower bound of problem $\mathcal{P}2$ is shrunk. Moreover, the strategic space of user grouping matrix $\bm{x}$ is finite. Thus, Algorithm \ref{alg1} is bound to stop in finite steps for any given $\delta>0$. The proof is stated in section 2.4 of \cite{Angeles2012}  in detail.\end{proof}

In Algorithm \ref{alg1}, we solve power allocation problem ${\mathcal{S}^{(k)}}$ by the interior point method, and its computational complexity is $O\big(N(NM)^3\big)$\cite{boyd2004convex}. If we solve the relaxed master problem $\mathcal{M}3$ by exhaustive search algorithm, the computational complexity will be unbearable. We introduce the way to solve the relaxed master problem $\mathcal{M}3$ in Section \ref{RelaxedMasterProblem}.

\section{Relaxed Master Problem Analysis and Solutions Based on Graph Theory}\label{RelaxedMasterProblem}

In this section, we convert the relaxed master user grouping problem into a problem of searching for some special negative loops in a graph composed of users based on graph theory. Two algorithms to find these loops are proposed. In order to find the way to reduce the values of $\mathcal{L}(\bm{q}^{(k_1)},\bm{\lambda}^{(k_1)}, \bm{x})$ and $\mathcal{L}'(\bm{q}^{(k_2)}, \bm{\nu}^{(k_2)}, \bm{x})$ in the relaxed master user grouping problem $\mathcal{M}3$, we introduce three definitions as follows:

\begin{def1}\label{defe1}
For $L$ users numbered by $n_1,n_2,\ldots,n_L$ in different groups, if problem $\mathcal{S}^{(k)}$ is feasible and $\mathcal{L}(\bm{q}^{(k)},\bm{\lambda}^{(k)}, \bm{x})$ can be reduced by $n_1\rightarrow n_2,\ldots,n_{L-2}\rightarrow n_{L-1}$ and putting user $n_{L-1}$ into the group of user $n_L$, or if problem $\mathcal{S}^{(k)}$ is infeasible, and $\mathcal{L}'(\bm{q}^{(k)},\bm{\nu}^{(k)}, \bm{x})$ can be reduced by $n_1\rightarrow n_2,\ldots,n_{L-2}\rightarrow n_{L-1}$ and putting user $n_{L-1}$ into the group of user $n_L$, these users compose a \emph{k-shift union}.
\end{def1}

To explain the meaning of $n\rightarrow n'$, we assume that with grouping strategy $\bm{x}$, user $n$ and user $n'$ are in group $g$ and $g'$, respectively. That is, $x_{gn}=1$, $x_{g'n'}=1$.
Then $n\rightarrow n'$ means transferring user $n$ into group $g'$ and removing user $n'$ from group $g'$, that is, $x_{gn}=0$, $x_{gn'}=1$, $x_{g'n'}=0$.

\begin{def1}\label{defe2}
For $L$ users numbered by $n_1,n_2,\ldots,n_L$ in different groups, if problem $\mathcal{S}^{(k)}$ is feasible and $\mathcal{L}(\bm{q}^{(k)},\bm{\lambda}^{(k)}, \bm{x})$ can be reduced by $n_1\rightarrow n_2,\ldots,n_L\rightarrow n_1$, or if problem $\mathcal{S}^{(k)}$ is infeasible, and $\mathcal{L}'(\bm{q}^{(k)},\bm{\nu}^{(k)}, \bm{x})$ can be reduced by $n_1\rightarrow n_2,\ldots,n_L\rightarrow n_1$, these users compose a \emph{k-exchange union}.
\end{def1}

\begin{def1}\label{defe3-}
For grouping strategy $\bm{x}$, if the value of $\xi$ in problem $\mathcal{M}3$ can not be reduced by any \emph{shift union} or \emph{exchange union} with all the constraints in problem $\mathcal{M}3$ satisfied, grouping matrix $\bm{x}$ is called \emph{all-stable solution}.
\end{def1}

To find the \emph{shift union}s and \emph{exchange union}s among users, we analyze the rules that values of $\mathcal{L}(\bm{q}^{(k)},\bm{\lambda}^{(k)}, \bm{x})$ and $\mathcal{L}'(\bm{q}^{(k)}, \bm{\nu}^{(k)}, \bm{x})$ change after changing user grouping matrix $\bm{x}$. By dividing the expressions of $\mathcal{L}(\bm{q}^{(k)},\bm{\lambda}^{(k)}, \bm{x})$ and $\mathcal{L}'(\bm{q}^{(k)}, \bm{\nu}^{(k)}, \bm{x})$, i.e., $(\ref{0016})$ and $(\ref{0019})$ in groups,
we define two weighted interference plus noise variables $\omega_g(\bm{x};\bm{q},\bm{\lambda})$ and $\omega'_g(\bm{x};\bm{q},\bm{\nu})$ as follows:

{\small
		\begin{equation}\label{g001}
\begin{aligned}
&\omega_g(\bm{x};\bm{q},\bm{\lambda})\\=&\sum\limits_{{j=1}}^{N}x_{gj}\lambda_j\left(\left({\sigma^2+\sum\limits_{i=1}^{N}x_{gi}\sum\limits_{m=1}^{M}q_{mi}^2{\upsilon_{gmji}}} \right)^{\tfrac{1}{2}} \gamma_j^{\tfrac{1}{2}}- {{\sum\limits_{m=1}^{M}q_{mj}x_{gj}}{\vartheta_{gmj}}}\right)\\
\end{aligned}
\end{equation}}
{\small
		\begin{equation}\label{g002}
\begin{aligned}
&\omega'_g(\bm{x};\bm{q},\bm{\nu})\\=&\sum\limits_{j=1}^{N}x_{gj}\nu_j\left(\left({\sigma^2+\sum\limits_{i=1}^{N}x_{gi}\sum\limits_{m=1}^{M}q_{mi}^2{\upsilon_{gmji}}} \right)^{\tfrac{1}{2}} \gamma_j^{\tfrac{1}{2}}- {{\sum\limits_{m=1}^{M}q_{mj}x_{gj}{\vartheta_{gmj}}}}\right)
\end{aligned}
		\end{equation}}
Combining $(\ref{0016})$, $(\ref{0019})$, $(\ref{g001})$ and $(\ref{g002})$, we have $\mathcal{L}(\bm{q}^{(k)},\bm{\lambda}^{(k)}, \bm{x})=\sum\limits_{g=1}^{G}\omega_g(\bm{x};\bm{q},\bm{\lambda})$ and $\mathcal{L}'(\bm{q}^{(k)}, \bm{\nu}^{(k)}, \bm{x})=\sum\limits_{g=1}^{G}\omega'_g(\bm{x};\bm{q},\bm{\nu})$. Next we investigate how user grouping strategy changing impacts the value of $\omega_g(\bm{x};\bm{q},\bm{\lambda})$ and $\omega'_g(\bm{x};\bm{q},\bm{\nu})$.

For convenience, we assume that in grouping strategy $\bm{x}$, user $n$ is in group $g_n$. Then we construct a directed graph $G(\mathcal{N}, \mathcal{E}^{(k)}; \bm{x})$, where $\mathcal{N}$ is the set of nodes composed of users and $\mathcal{E}^{(k)}$ is the set of edges existing between two users in different groups. The adjacency matrix of graph $G(\mathcal{N}, \mathcal{E}^{(k)}; \bm{x})$ is denoted by $\bm{a}^{(k)}$, and we set that:
{\small
\begin{equation}\label{g007}
a_{ij}^{(k)} =\left\{ \begin{array}{ll}
\omega_{g_j}(x_{g_ji}=1,x_{g_jj}=0, \bm{x}_{-i,j}&;\bm{q}^{(k)},\bm{\lambda}^{(k)})-\omega_{g_j}(\bm{x};\bm{q}^{(k)},\bm{\lambda}^{(k)})
\\ &\text{if $g_i\neq g_j$, $\mathcal{S}^{(k)}$ is feasible}\\
 \omega'_{g_j}(x_{g_ji}=1,x_{g_jj}=0, \bm{x}_{-i,j}&;\bm{q}^{(k)},\bm{\nu}^{(k)})-\omega'_{g_j}(\bm{x};\bm{q}^{(k)},\bm{\nu}^{(k)})
\\ &\text{if $g_i\neq g_j$, $\mathcal{S}^{(k)}$ is infeasible}\\
\infty, &\text{if $g_i=g_j$}
 \end{array}\right.
\end{equation}}where $\bm{x}_{-i,j}$ denotes the user grouping strategies of users except users $i$ and $j$, thus $(x_{g_ji}=1,x_{g_jj}=0, \bm{x}_{-i,j})$ represents that user $i$ is in group $g_j$, user $j$ is not in group $g_j$, and the user grouping strategies of users except users $i$ and $j$ consistent with $\bm{x}$.
The relation between problem $\mathcal{M}3$ and graph $G(\mathcal{N}, \mathcal{E}^{(k)}; \bm{x})$ can be indicated with the following propositions.

\begin{pro1}
\label{pro3}

For $L$ users numbered by $n_1,n_2,\ldots,n_L$ in different groups, if problem $\mathcal{S}^{(k)}$ is feasible and $\mathcal{L}(\bm{q}^{(k)},\bm{\lambda}^{(k)}, \bm{x})$ can be reduced by $n_1\rightarrow n_2,\ldots,n_L\rightarrow n_1$, these users can compose a negative loop $n_1- >n_2->\ldots->n_L->n_1$ in graph $G(\mathcal{N}, \mathcal{E}^{(k)}; \bm{x})$.

\end{pro1}

\begin{proof}

We assume that the user grouping matrix before and after $n_1\rightarrow n_2,\ldots,n_L\rightarrow n_1$ are $\bm{x}$ and $\bm{x}'$. Then the difference of $\mathcal{L}(\bm{q}^{(k)},\bm{\lambda}^{(k)}, \bm{x})$ is{\small
		\begin{equation}\label{g008}
\begin{aligned}
\nabla_{\mathcal{L}}=&\mathcal{L}(\bm{q}^{(k)},\bm{\lambda}^{(k)}, \bm{x}')-\mathcal{L}(\bm{q}^{(k)},\bm{\lambda}^{(k)}, \bm{x})\\
=&\sum\limits_{n=1}^{N}\lambda_n^{(k)}\left(\left({\sigma^2+\sum\limits_{g=1}^{G}x'_{gn}\sum\limits_{i=1}^{N}x'_{gi}\sum\limits_{m=1}^{M}(q_{mi}^{(k)})^2{\upsilon^{(k)}_{gmni}}} \right)^{\tfrac{1}{2}} \gamma_n^{\tfrac{1}{2}}\right.\\
&\left.- {{\sum\limits_{m=1}^{M}q_{mn}\sum\limits_{g=1}^{G}x_{gn}{\vartheta^{(k)}_{gmn}}}}\right)\\&-\sum\limits_{n=1}^{N}\lambda_n^{(k)}\left(\left({\sigma^2+\sum\limits_{g=1}^{G}x_{gn}\sum\limits_{i=1}^{N}x_{gi}\sum\limits_{m=1}^{M}(q_{mi}^{(k)})^2{\upsilon^{(k)}_{gmni}}} \right)^{\tfrac{1}{2}} \gamma_n^{\tfrac{1}{2}}\right.\\
&\left.- {{\sum\limits_{m=1}^{M}q_{mn}\sum\limits_{g=1}^{G}x_{gn}{\vartheta^{(k)}_{gmn}}}}\right)
\end{aligned}
\end{equation}}
We assume that users $n_1,n_2,\ldots,n_L$ are in groups $g_1,g_2,\ldots,g_L$, respectively, note that the value of
$\lambda_n^{(k)}\left(\left({\sigma^2+\sum\limits_{g=1}^{G}x_{gn}\sum\limits_{i=1}^{N}x_{gi}\sum\limits_{m=1}^{M}(q_{mi}^{(k)})^2{\upsilon^{(k)}_{gmni}}} \right)^{\tfrac{1}{2}} \gamma_n^{\tfrac{1}{2}}\right.$ $-\left. {{\sum\limits_{m=1}^{M}q_{mn}\sum\limits_{g=1}^{G}x_{gn}{\vartheta^{(k)}_{gmn}}}}\right),~~n\notin\{n_1,n_2,\ldots,n_L\}$ will not change after $n_1\rightarrow n_2,\ldots,n_L\rightarrow n_1$. Then according to (\ref{g008}), we have:{\small
		\begin{equation}\label{g009}
\begin{aligned}
\nabla_{\mathcal{L}}=&\sum\limits_{l=1}^{L}\sum\limits_{n=1}^{N}x'_{g_ln}\lambda_n^{(k)}\left(\left({\sigma^2+\sum\limits_{i=1}^{N}x'_{g_li}\sum\limits_{m=1}^{M}(q_{mi}^{(k)})^2{\upsilon^{(k)}_{gmni}}} \right)^{\tfrac{1}{2}} \gamma_n^{\tfrac{1}{2}} -\right.\\
&\left. {{\sum\limits_{m=1}^{M}q_{mn}\sum\limits_{g=1}^{G}x_{gn}{\vartheta^{(k)}_{gmn}}}}\right)\\&-\sum\limits_{l=1}^{L}\sum\limits_{n=1}^{N}x_{g_ln}\lambda_n^{(k)}\left(\left({\sigma^2+\sum\limits_{i=1}^{N}x_{g_li}\sum\limits_{m=1}^{M}(q_{mi}^{(k)})^2{\upsilon^{(k)}_{gmni}}} \right)^{\tfrac{1}{2}} \gamma_n^{\tfrac{1}{2}}-\right.\\
&\left. {{\sum\limits_{m=1}^{M}q_{mn}\sum\limits_{g=1}^{G}x_{gn}{\vartheta^{(k)}_{gmn}}}}\right)\\
=&\sum\limits_{l=1}^{L}a_{lj}^{(k)}
\end{aligned}
\end{equation}}
where $j=mod(l,L)+1$.

Then the proof of Proposition \ref{pro3} is concluded.\end{proof}

\begin{pro1}
\label{pro4}

For $L$ users numbered by $n_1,n_2,\ldots,n_L$ in different groups, if problem $\mathcal{S}^{(k)}$ is infeasible, and $\mathcal{L}'(\bm{q}^{(k)},\bm{\nu}^{(k)}, \bm{x})$ can be reduced by $n_1\rightarrow n_2,\ldots,n_L\rightarrow n_1$, these users can compose a negative loop $n_1- >n_2->\ldots->n_L->n_1$ in graph $G(\mathcal{N}, \mathcal{E}^{(k)}; \bm{x})$.\end{pro1}

\begin{proof}

The proof is similar to Proposition \ref{pro3}.\end{proof}

According to Proposition \ref{pro3} and Proposition \ref{pro4}, we can find the grouping changing method to reduce the value of $\mathcal{L}(\bm{q}^{(k)},\bm{\lambda}^{(k)}, \bm{x})$ or $\mathcal{L}'(\bm{q}^{(k)},\bm{\nu}^{(k)}, \bm{x})$ by searching for the negative loop with all users in different groups, we call them negative differ-group loop. However, the number of users in each group will not change after $n_1\rightarrow n_2,\ldots,n_L\rightarrow n_1$. To find the grouping changing method that can lead to arbitrary number of users in each group and reduce the value of $\mathcal{L}(\bm{q}^{(k)},\bm{\lambda}^{(k)}, \bm{x})$ or $\mathcal{L}'(\bm{q}^{(k)},\bm{\nu}^{(k)}, \bm{x})$, we expand graph $G(\mathcal{N}, \mathcal{E}^{(k)}; \bm{x})$ to  $G(\mathcal{N}^e, \mathcal{E}^{(k)}; \bm{x})$ by adding a virtual user to each group. The virtual user added to group $g$ is numbered by $n_g^v$. We set the achievable SINR of these virtual users to $0$, so these virtual users will not be allocated any power.

\begin{pro1}
\label{pro_convert}If users $n_1,n_2,\ldots,n_{L-1},n_L$ compose a \emph{shift union} with grouping strategy $\bm{x}$, users $n_1,n_2,\ldots,n_{L-1},n_{g_L}^v$ can compose an \emph{exchange union} with grouping strategy $\bm{x}$.\end{pro1}

\begin{proof}
Obviously, if grouping strategy $\bm{x}$ is changed to $\bm{x}'$ after $n_1\rightarrow n_2,\ldots,n_{L-2}\rightarrow n_{L-1}$ and putting user $n_{L-1}$ into the group of user $n_L$, it will also be changed to $\bm{x}'$ after $n_1\rightarrow n_2,\ldots,n_{L-2}\rightarrow n_{L-1},n_{L-1}\rightarrow n_{g_L}^v$. So the proof of Proposition \ref{pro_convert} is concluded.\end{proof}

\begin{thm1}
\label{thm_all}
For grouping strategy $\bm{x}$, if the value of $\xi$ in problem $\mathcal{M}3$ can not be reduced by any negative differ-group loop in graph $G(\mathcal{N}^e, \mathcal{E}^{(k)}; \bm{x})$ with all the constraints in problem $\mathcal{M}3$ satisfied, grouping matrix $\bm{x}$ is called \emph{all-stable solution}.
\end{thm1}

\begin{proof}
According to Proposition \ref{pro3}, Proposition \ref{pro4} and Proposition \ref{pro_convert}, if there is a \emph{shift union} or an \emph{exchange union} among all real users and virtual users, there must be a negative differ-group loop in graph $G(\mathcal{N}^e, \mathcal{E}^{(k)}; \bm{x})$. Therefore, if the value of $\xi$ in problem $\mathcal{M}3$ can not be reduced by any negative differ-group loop in graph $G(\mathcal{N}^e, \mathcal{E}^{(k)}; \bm{x})$ with all the constraints in problem $\mathcal{M}3$ satisfied, the value of $\xi$ in problem $\mathcal{M}3$ can not be reduced by any \emph{shift union} or \emph{exchange union} with all the constraints in problem $\mathcal{M}3$ satisfied, i.e., the grouping matrix $\bm{x}$ is \emph{all-stable solution}.
\end{proof}

{
\setlength{\abovecaptionskip}{-0.02cm}
\begin{figure}[ht]
	\centering
	\includegraphics[width=0.48\textwidth]{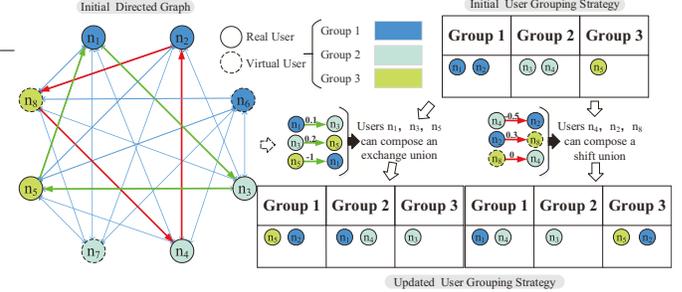}
	\caption{Illustration of the directed graph with ``shift union'' and ``exchange union''.}
	\label{fig:digraph}
\end{figure}}

{In order to explain the concepts of \emph{shift union} and \emph{exchange union} more clearly, an illustration of the directed graph composed of 8 nodes is shown in Fig.{\ref{fig:digraph}} with edges among users in different groups, where the real users and virtual users are represented by solid circles and dotted circles, respectively. The users are divided into three groups, and the users in the same colour are grouped into the same group with initial user grouping strategy. In this directed graph, two negative differ-group loops $n_1- >n_3->n_5->n_1$ and $n_4- >n_2->n_8->n_4$ are found. Among them, users $n_1$, $n_3$, $n_5$ can compose an \emph{exchange union}, and users $n_1$, $n_3$, $n_5$ can compose a \emph{shift union}. Then, $\mathcal{L}(\bm{q}^{(k)},\bm{\lambda}^{(k)}, \bm{x})$ (if problem $\mathcal{S}^{(k)}$ is infeasible) or $\mathcal{L}'(\bm{q}^{(k)},\bm{\nu}^{(k)}, \bm{x})$ (if problem $\mathcal{S}^{(k)}$ is infeasible) can be reduced by $n_1\rightarrow n_3, n_3\rightarrow n_5, n_5\rightarrow n_1$ or $n_4\rightarrow n_2, n_2\rightarrow n_8, n_8\rightarrow n_4$. It is worth noting that the initial user grouping strategy is an \emph{all-stable solution}, if there are no negative differ-group loop can be found in the initial directed graph.
}

\begin{algorithm}[htbp]\label{alg2}\small
\setlength\abovedisplayskip{1pt}
\setlength\belowdisplayskip{0pt}
 \caption{Graph Theory Based Algorithm to Solve Master Problem (GBMA)}\label{alg:2}
  \KwIn{Relaxed Master Problem $\mathcal{M}3$, Grouping Matrix $\bm{x}^{(k)}$, }

  \KwOut{Grouping Matrix $\bm{x}^{(k+1)}$}
  Create set of infeasible loops $\mathcal{A}=\varnothing$;

   \Repeat{$maxL<0$}{
        Find $k_{max}=arg\max\limits_{i\leq k, \mathcal{S}^{(i)} \text{is infeasible.}}~~\mathcal{L}'(\bm{q}^{(i)},\bm{\lambda}^{(i)}, \bm{x}^{(k)})$;

        $maxL=\mathcal{L}'(\bm{q}^{(k_{max})},\bm{\lambda}^{(k_{max})}, \bm{x}^{(k)})$;

        \If{$maxL>0$.}{
        Create graph $G(\mathcal{N}^e, \mathcal{E}^{(k_{max})}; \bm{x}^{(k_{max})})$;

        Search for negative differ-group loop $\bm{L}^{(k_{max})}\notin \mathcal{A}$ in graph $G(\mathcal{N}^e, \mathcal{E}^{(k)}; \bm{x}^{(k)})$;

        Change grouping matrix $\bm{x}^{(k)}$ to $\bm{x}^{(k+1)}$ according to loop $\bm{L}^{(k)}$ based on Proposition \ref{pro3}-\ref{pro4}.

         \For{$i=1 \textrm{ to }k-1$}{

        \If{$\mathcal{S}^{(i)}$ is infeasible.}{

         \If{$\mathcal{L}'(\bm{q}^{(k)},\bm{\nu}^{(k)}, \bm{x}^{(k+1)})> maxL$}{

        Add loop $\bm{L}^{(k_{max})}$ to $\mathcal{A}$;
                 Jump to step 6;
        }{}
        }

        }
$\bm{x}^{(k)}$ = $\bm{x}^{(k+1)}$;$\mathcal{A}=\varnothing$;
        }

        }

  \Repeat{Cannot find an appropriate negative differ-group loop}{

  $\mathcal{A}=\varnothing$;

         Find $k_{max}=arg\max\limits_{i\leq k, \mathcal{S}^{(i)} \text{is feasible.}}~~\mathcal{L}(\bm{q}^{(i)},\bm{\lambda}^{(i)}, \bm{x}^{(k)})$;

         Create graph $G(\mathcal{N}^e, \mathcal{E}^{(k_{max})}; \bm{x}^{(k)})$;

        Search for negative differ-group loop $\bm{L}^{(k_{max})}\notin \mathcal{A}$ in graph $G(\mathcal{N}^e, \mathcal{E}^{(k_{max})}; \bm{x}^{(k)})$;

        Change grouping matrix $\bm{x}^{(k)}$ to $\bm{x}^{(k+1)}$ according to loop $\bm{L}^{(k)}$ based on Proposition \ref{pro3}-\ref{pro4}.

        \For{$i=1 \textrm{ to }k$}{

        \eIf{$\mathcal{S}^{(i)}$ is feasible.}{

         \If{$\mathcal{L}(\bm{q}^{(i)},\bm{\lambda}^{(i)}, \bm{x}^{(k+1)})>\mathcal{L}(\bm{q}^{(i)},\bm{\lambda}^{(i)}, \bm{x}^{(k_{max})})$}{

        Add loop $\bm{L}^{(k_{max})}$ to $\mathcal{A}$;
                 Jump to step 22;
         }
        }
        {

         \If{$\mathcal{L}'(\bm{q}^{(k)},\bm{\nu}^{(k)}, \bm{x}^{(k+1)})>0$}{

        Add loop $\bm{L}^{(k_{max})}$ to $\mathcal{A}$;
                 Jump to step 22;
        }

        }
        }

$\bm{x}^{(k)}=\bm{x}^{(k+1)}$;
}

Return $\bm{x}^{(k+1)}$.
\end{algorithm}

Graph theory based algorithm to solve relaxed master problem $\mathcal{M}3$ is shown in Algorithm \ref{alg2}. In this algorithm, we first search for a new user grouping matrix which satisfies all the infeasi-constraints in problem $\mathcal{M}3$ in steps 2-17. In each iteration of steps 3-16, we change the user grouping matrix to reduce the value of  $\max\limits_{i\leq k, \mathcal{S}^{(i)} \text{is infeasible.}}\mathcal{L}'(\bm{q}^{(i)},\bm{\lambda}^{(i)}, \bm{x}^{(k)})$, where $\mathcal{A}$ is the set of the loops according to which we can not reduce the value of  $\max\limits_{i\leq k, \mathcal{S}^{(i)} \text{is infeasible.}}\mathcal{L}'(\bm{q}^{(i)},\bm{\lambda}^{(i)}, \bm{x}^{(k)})$. Then we search for the solution of problem $\mathcal{M}3$ in steps 18-36. In each iteration of steps 19-35, we change the user grouping matrix to reduce the value of  $\max\limits_{i\leq k, \mathcal{S}^{(i)} \text{is feasible.}}\mathcal{L}(\bm{q}^{(i)},\bm{\lambda}^{(i)}, \bm{x}^{(k)})$, where $\mathcal{A}$ is the set of the loops according to which we can not reduce the value of  $\max\limits_{i\leq k, \mathcal{S}^{(i)} \text{is feasible.}}\mathcal{L}(\bm{q}^{(i)},\bm{\lambda}^{(i)}, \bm{x}^{(k)})$.

\begin{cor1}\label{Cor_converge}
Algorithm \ref{alg2} can converge to \emph{all-stable solution} in finite iterations.
\end{cor1}

\begin{proof}
The number of nodes in the graph is limited by the numbers of users and groups, so the number of the negative differ-group loops in the graph is infinite. In addition, the optimal value of problem $\mathcal{M}3$ in each iteration of Algorithm \ref{alg2} will descend and the feasibility of the outputted user grouping matrix for problem $\mathcal{M}3$ in each iteration can be guaranteed by step 26. Therefore, Algorithm \ref{alg2} will stop after finite iterations and output a solution without negative differ-group loop, i.e., \emph{all-stable solution}.
\qedhere
\end{proof}

\begin{algorithm}[htbp]\label{alg2+}\small
\setlength\abovedisplayskip{1pt}
\setlength\belowdisplayskip{0pt}
 \caption{Extended Bellman-Ford Algorithm to Search for Negative Differ-group Loops in GBMA (EBSA)}\label{alg:2}
  \KwIn{Group $G(\mathcal{N}^e, \mathcal{E}^{(k_{max})}; \bm{x}^{(k)})$, Set of infeasible loops $\mathcal{A}$, Adjacency matrix $\bm{a}^{(k)}$}
  \KwOut{\emph{Negative differ-group loop} $\bm{L}^{(k)}$}
  Create super node.

\For{$j=1 \textrm{ to }(G+N)$}{
$w_n=0$, $\mathcal{T}_n=\varnothing$, $n\in \mathcal{V}$;
}

  \Repeat{$\mathcal{T}_n$, $ n\in\mathcal{V}$ do not change}{
        \For{$i=1 \textrm{ to }(G+N)$}{
        \For{$j=1 \textrm{ to }(G+N)$}{
        \If{$(w_{j}>w_{i}+a^{(k)}_{ij}) \&  (a^{(k)}_{kj}\neq \infty, \forall k\in\mathcal{T}_i\setminus \{j\})$}{

        \If{There is a loop $\bm{L}$ in $(\mathcal{T}_i\cup \{i\})$ and $\bm{L}\notin \mathcal{A}$}{

         $\mathcal{T}_j=\mathcal{T}_i\cup \{i\}$, $w_{j}= w_{i}+a^{(k)}_{ij}$
         }
        }
        \If{$(w_{j}>w_{i}+a^{(k)}_{ij}) \& (\exists a^{(k)}_{kj}= \infty,k\in\mathcal{T}_i\setminus \{j\})$}{
         Find the shortest path $\mathcal{T}'_i$ from the super node to user $i$ with $ a^{(k)}_{kj}\neq \infty, \forall k\in\mathcal{T}'_i\setminus \{j\}$, assume the distance of path $\mathcal{T}'_i$ is $m'_{i}$;

        \If{$w_{j}>m'_{i}+a^{(k)}_{ij}$}{
        \If{There is a loop $\bm{L}$ in $(\mathcal{T}'_i\cup \{i\})$ and $\bm{L}\notin \mathcal{A}$}{
         $\mathcal{T}_j=\mathcal{T}'_i\cup \{i\}$, $w_{j}= m'_{i}+a^{(k)}_{ij}$
         }
		}
        }
        \If{$\|\mathcal{T}_j\|> G$}{
         Find the \emph{negative differ-group loop} $\bm{L}$ in $\mathcal{T}_j$;
         \textbf{return} $\bm{L}$ and break;
		}

		}

		}
          }

\end{algorithm}

To find these negative differ-group loops, we extend the Bellman-Ford algorithm to Algorithm \ref{alg2+} \cite{guo2021interference} \cite{7763835} \cite{Maccari2015}, where the negative differ-group loops in the set of infeasible loops $\mathcal{A}$ are avoided to be outputted. In this algorithm, we first create a super node. The distance from the super node to node $n$ is set to $w_n=0$, and the path from super node to node $n$, $\mathcal{T}_n$, $n\in \mathcal{V}$, is initialized in steps 2-4. Then these paths are constantly relaxed in steps 5-26. In each step of relaxing, the users in the same group are avoided to be added to the same path from super node to any node as step 8 and step 13 show. And the loops in $\mathcal{A}$ are avoided to form in any path $\mathcal{T}_n$ as step 9 and step 16 show. Therefore, according to the principle of Bellman-Ford algorithm, if the path from super node to any node is longer than the number of groups, there must be a negative differ-group loop in this path \cite{7287776}.

To obtain an appropriate solution of problem $\mathcal{P}1$ with polynomial time, we also design a greedy fast algorithm for the solving of problem $\mathcal{M}3$ as shown in Algorithm \ref{alg3}. This searching algorithm starts from the minimal edge in the graph. Then we iteratively search for the minimal output edge until there exists no next output edge. In this process, the set of loops that have been rejected in steps 5-7 of Algorithm \ref{alg3} is avoided to be outputted.

\begin{algorithm}[htbp]\label{alg3}\small
\setlength\abovedisplayskip{1pt}
\setlength\belowdisplayskip{0pt}
\setlength{\abovecaptionskip}{-0.02cm}
\setlength{\belowcaptionskip}{-0.1cm}
 \caption{Greedy Fast Algorithm to Search for Negative Differ-group Loop in GBMA (GFSA)}\label{alg:2}
  \KwIn{Group $G(\mathcal{N}^e, \mathcal{E}^{(k_{max})}; \bm{x}^{(k)})$, Set of infeasible loops $\mathcal{A}$, Adjacency matrix $\bm{a}^{(k)}$}
  \KwOut{\emph{Negative differ-group loop} $\bm{L}^{(k)}$}

        \For{$t=1 \textrm{ to }(N+G)$}{
        $\mathcal{T}=\varnothing$;

         Find the minimal edge $a^{(k)}_{ij}=\min\limits\{a^{(k)}_{ij}|i,j\in\mathcal{N}^e\}$;

        $\mathcal{T}=\mathcal{T}\cup \{i\}\cup \{j\}$, $m_t= a^{(k)}_{ij}$;

        \If{$m_t+a_{ji}^{(k)}<0$\&$\mathcal{T}\notin \mathcal{A}$}{
        $\bm{L}^{(k)}\leftarrow\mathcal{T}$;
        \textbf{return} $\bm{L}^{(k)}$.
        }

        $a^{(k)}_{ij}=\infty$, $i= j$;
        \For{$l=3 \textrm{ to }G$}{
        Find the minimal output edge of node $i$ : $a^{(k)}_{ij}=\min\limits\{a^{(k)}_{ij}|j\notin\mathcal{T} \}$;

        $m_t= m_t+a^{(k)}_{ij}$, $\mathcal{T}=\mathcal{T}\cup \{j\}$;

        \If{$m_t+a_{ji}^{(k)}<0$\&$\mathcal{T}\notin \mathcal{A}$}{
        $\bm{L}^{(k)}\leftarrow\mathcal{T}$;
        \textbf{return} $\bm{L}^{(k)}$.
        }
        $i= j$;
        }

        }

\end{algorithm}\emph{Computational Complexity Analysis:} In GFSA, complexity of steps $ 10 $-$ 15 $ is $O\big(G+N\big)$, and complexity of steps $ 2 $-$ 8 $ is $O\big((G+N)^2\big)$. Hence, computational complexity of GFSA is $O\big((G+N)^3\big)$. Assume that GFSA is repeated $C$ times in Algorithm \ref{alg2}. Then apparently computational complexity of GBMA is $O\big(C(G+N)^3\big)$.
We solve power allocation problem ${\mathcal{S}^{(k)}}$ by the interior point method, and its computational complexity is $O\big(N(NM)^3\big)$\cite{boyd2004convex}. Then computational complexity of the proposed fast greedy algorithm is $O\big(max(N^2(NM)^3,CN(G+N)^3)\big)$.

\section{Performance Evaluation}\label{PerformanceEvaluation}

In this section, we evaluate the performance of the proposed two algorithms in cell-free massive MIMO systems in terms of transmit power, {interference from the desired
signals of the other users} in the same group, etc. In the simulations, APs and users are randomly placed in a 3km$\times$3km rectangular area. We set large-scale channel gain to $128.1+37.6log_{10}$ ($d_n$ [km]) dB. The small-scale fading follows an i.i.d. Gaussian distribution.
Some default values in the simulations are shown in Table \ref{table2} \cite{8756265}.

\begin{table}
\setlength{\abovecaptionskip}{-0.04cm}
\setlength\abovedisplayskip{1pt}
\setlength\belowdisplayskip{1pt}
\fontsize{8}{12}\selectfont
	\centering\label{table2}
	\caption{Main Notations}
	\begin{tabular} {|m{150pt}|	m{50pt}|}
    \hline
		Parameter		& Value						 	\\
    \hline
		Number of users(N)				& 200 										\\
    \hline
        Number of groups(G)				& 5 										\\	
    \hline
		Number of APs(M)				& 200 										\\
    \hline
		 Bandwidth(B) 		& 20 MHz										\\
    \hline
		 Noise power spectral density($N_0$) 		& -174 dBm/Hz									\\	
    \hline		
		 Power of pilot signal ($ \rho_r$) 		& 200 mW 									\\		
    \hline
		Length of pilot sequences ($\tau$) 		& $2\lceil N/G\rceil$							\\	
    \hline
		Target data rate  		& 0.1-1.5 Mbps							\\
    \hline
	\end{tabular}
	\label{table1}
\end{table}

\begin{figure}[htb]
\setlength{\belowcaptionskip}{-0.5cm}
\setlength{\belowdisplayskip}{-1.2cm}
\centering
\subfigure[]{
\label{fig:dd}
\includegraphics[height=0.18\textwidth]{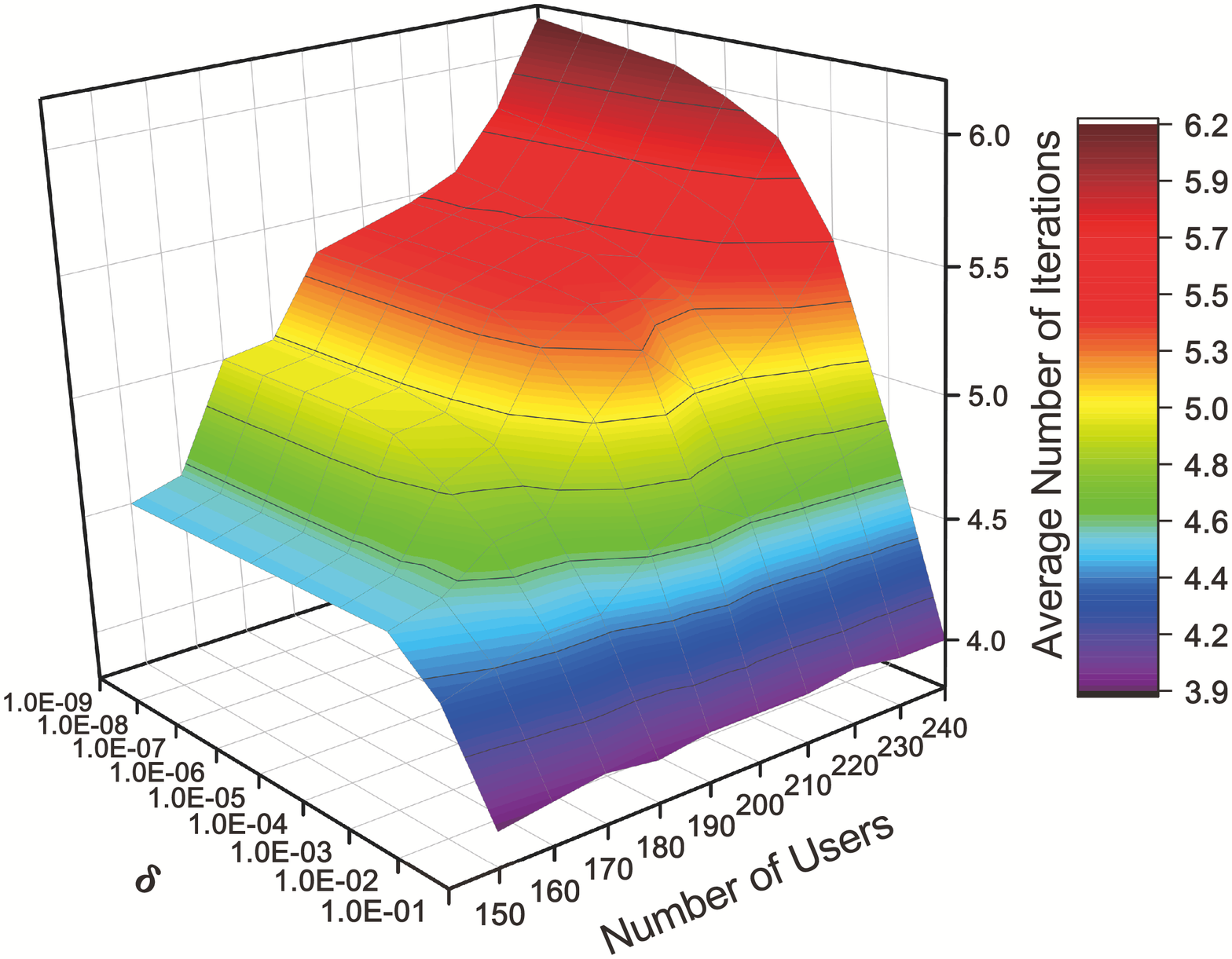}}
\subfigure[]{
\label{fig:sl}
\includegraphics[height=0.18\textwidth]{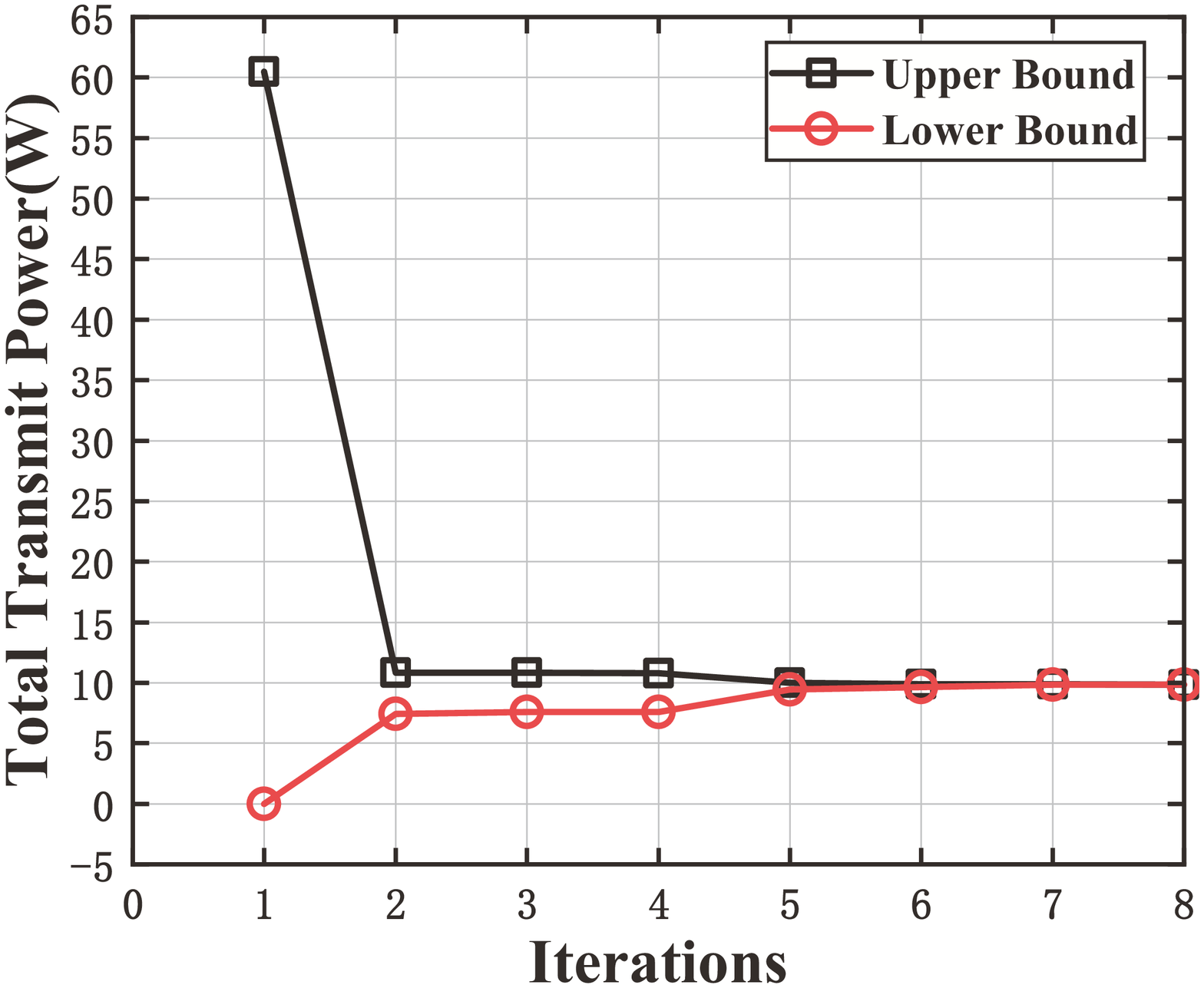}}
\centering
\caption{Performance comparison of different user grouping algorithms: a. Average number of iterations with different numbers of users and different $\delta$; b. Convergence of lower and upper bounds with the proposed fast greedy user grouping algorithm(200 users and 200 APs in total).}.
\label{fig:ddsl}
\end{figure}

\subsection{Convergence Performance}

To evaluate the convergence performance of the proposed fast greedy algorithm, in Fig. \ref{fig:dd}, we show the average number of iterations (denoted by $T_{iter}$) of steps 7-18 in Algorithm \ref{alg1} with different numbers of users and different $\delta$. We can see that $T_{iter}$ increases as number of users increases and $\delta$ reduces. In addition, $T_{iter}<10$ even when $N=240$ and $\delta=10^{-9}$. In Fig. \ref{fig:sl}, we show the process that the gap between the upper bound and lower bound of problem $\mathcal{P}2$ reduces. The results show that the proposed fast greedy algorithm converges rapidly, which illustrate the practicability of the proposed fast greedy user grouping algorithm. {As shown in Fig. \ref{fig:sl}, the gap between the upper bound and the lower bound shrinks as the number of iterations increases. According to the principle of benders decomposition, the optimal user grouping and power allocation strategy can be found if this gap $\delta$ is 0. The number of users is finite, hence the grouping strategy profile is finite. Every time we change the user grouping strategy, the total transmit power will not increase, which means that a grouping strategy profile will not be selected repeatedly. Thus we can obtain the optimal user grouping and power allocation strategy with a small enough $\delta$ in Algorithm 1.}

\subsection{Impacts of pilot signal}

As mentioned in section \ref{sec:02A}, channel estimation is carried out after user grouping in cell-free massive MIMO systems. Since the accuracy of channel estimation will affect the performance of cell-free massive MIMO with beamforming, we investigate the impacts of pilot signal on $\alpha$ in this subsection. $\alpha$ is the variance of MMSE estimate of channel fading ${h}$ as stated in (\ref{0000-1+1}). In Fig. \ref{fig:51}, we change the power of pilot signal $\rho_r$ and the length of pilot sequences $\tau_g$ to show its influence on $\alpha$ under the proposed user grouping algorithm, where $U_g$ means the of number of users in group $g$. {Considering the case of pilot reuse to observe the effects of non-orthogonal pilots, the length of pilot signal will be less than the number of users, i.e., $\tau_g<\sum _{i=1}^{N}x_{gi}$ \cite{Papazafeiropoulos2020}.
To investigate the effect of pilot signal with $\tau_g<\sum _{i=1}^{N}x_{gi}$, in this figure, the range of $\tau_g$ is set to $\tfrac{1}{2}U_g$, $\tfrac{1}{3}U_g$ and $\tfrac{1}{4}U_g$.} We can see that in Fig. \ref{fig:51}, as the increase of $\rho_r$ and $\tau_g$, the mean value of $\alpha$ will also increase, which agrees with (\ref{0000-1+1+1}). Then the channel estimation error is reduced as stated in (\ref{0000-1+3}). That is to say, by adding the length of pilot sequences or the power of the pilot signal, the accuracy of channel estimation can be increased. By user grouping, the number of users served by each time-slot can be reduced. Therefore, the length of pilot sequences to maintain the accuracy of channel estimation can be reduced by user grouping.

{
\vspace{-0.2cm}
\setlength{\abovecaptionskip}{-0.02cm}
\begin{figure}[ht]
	\centering
	\includegraphics[scale=0.28]{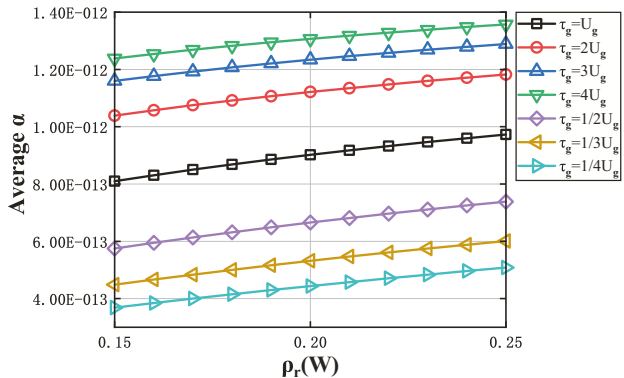}
	\caption{Influence of $\rho_r$ and $\tau$ on $\alpha _{mn}$.}
	\label{fig:51}
\end{figure}}

\begin{figure*}[htb]
\setlength{\belowcaptionskip}{-0.5cm}
\setlength{\belowdisplayskip}{-1.2cm}
\centering
\subfigure[]{
\label{fig:Pu}
\includegraphics[height=0.26\textwidth]{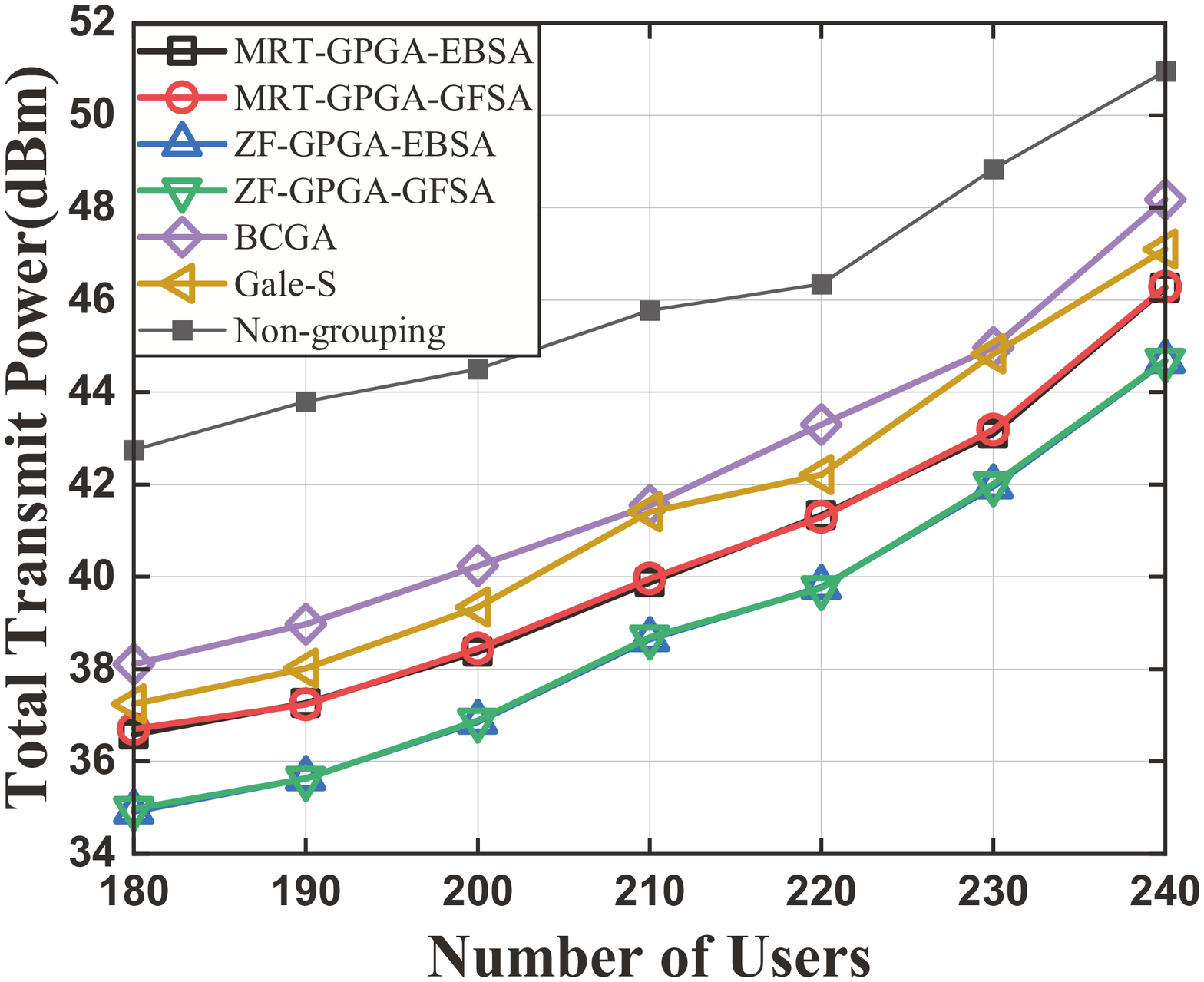}}
\subfigure[]{
\label{fig:PB}
\includegraphics[height=0.26\textwidth]{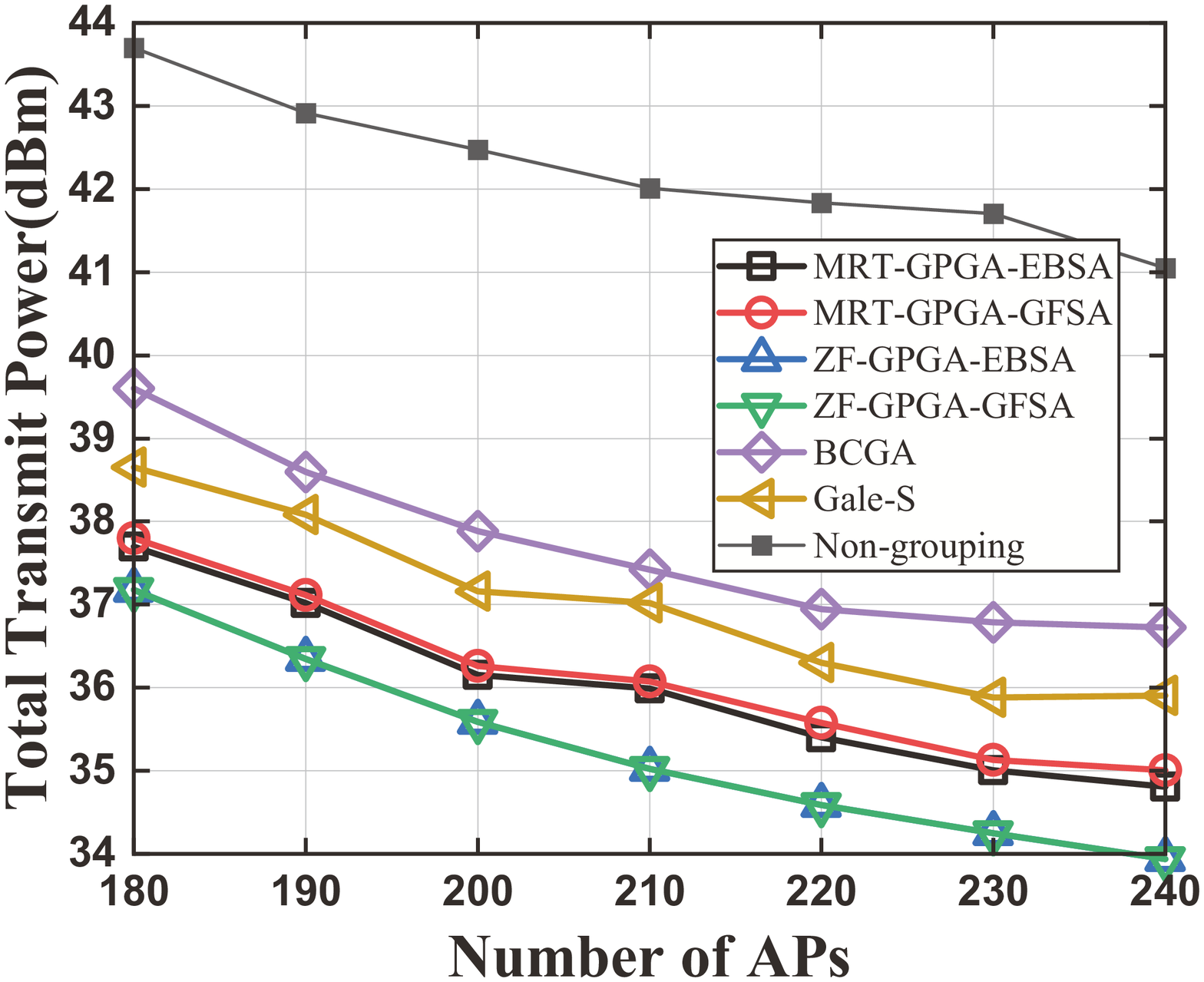}}
\subfigure[]{
\label{fig:rr}
\includegraphics[height=0.26\textwidth]{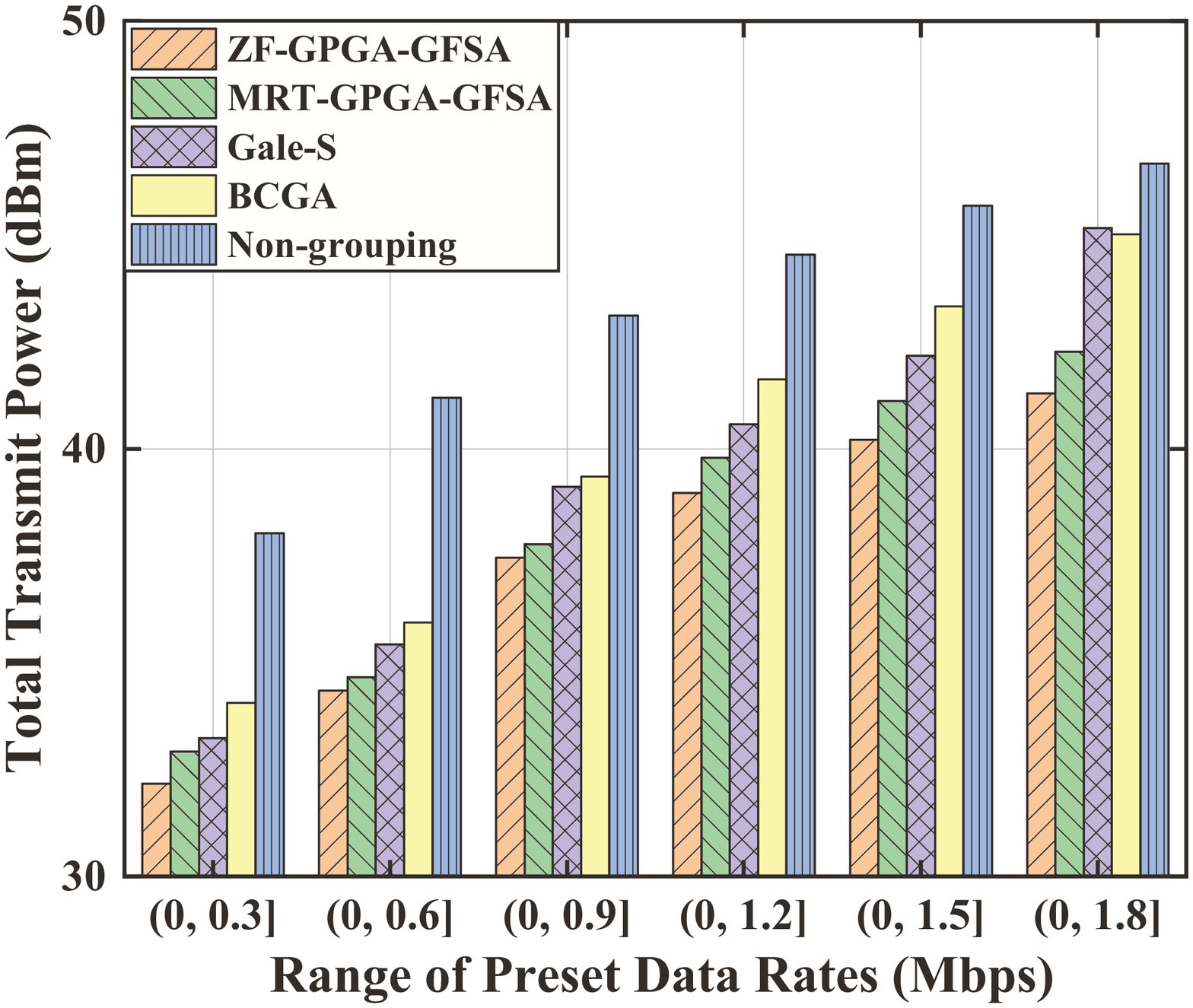}}
\centering
\caption{Performance comparison of different user grouping algorithms: a. Total transmit power vs. number of APs; b. Total transmit power vs. Number of APs; c. Total transmit power vs. Range of preset data rates (Mbps)}.
\label{fig:BS}
\end{figure*}

\subsection{Performance comparison}

{In this subsection, we compare four proposed user grouping algorithms (named ``MRT-GPGA-EBSA'', ``MRT-GPGA-GFSA'', ``ZF-GPGA-EBSA'' and ``ZF-GPGA-GFSA'', respectively) with the basic random user grouping algorithm(BCGA) and Gale-Shapley algorithm(Gale-S) \cite{Zhao2017ANon}, where each user prefers the group where interference is less and each group prefers to reject the access requests of the users with the highest requirements on power. The number of users in each group with Gale-S is equal.} To evaluate interference that users suffer, we define a mean-interference variable $I$ as follows:
{\small
		\begin{equation}\label{0061}
\begin{aligned}
I=&\tfrac{1}{N}\sum_{n=1}^{N}\bigg({\sum\limits_{g=1}^{G}x_{gn}\sum\limits_{i=1}^{N}x_{gi}\sum\limits_{m=1}^{M}p_{mi}\beta_{mn}\alpha_{gmi}}\bigg)
\end{aligned}
		\end{equation}}where $\bigg({\sum\limits_{g=1}^{G}x_{gn}\sum\limits_{i=1}^{N}x_{gi}\sum\limits_{m=1}^{M}p_{mi}\beta_{mn}\alpha_{gmi}}\bigg)$ is the right of denominator of $\mathrm{SINR}_n$ in (\ref{0001}).

We first vary the number of users and APs to show mean-interference variable $I$ with four different user grouping algorithms in Fig. \ref{fig:Iu} and Fig. \ref{fig:IB}, respectively.
The number of users increases from 150 to 240 with 200 APs and five groups in total in Fig. \ref{fig:Iu}, and the number of APs increases from 150 to 240 with 200 users and five groups in total in Fig. \ref{fig:IB}. We can see that mean-interference increases with the increase of users and the reduction of APs. The reason is that as the number of groups is given, the number of users sharing the same time-slot will increase with the increase of users. Then the value of ${\sum\limits_{g=1}^{G}x_{gn}\sum\limits_{i=1}^{N}x_{gi}\sum\limits_{m=1}^{M}p_{mi}\beta_{mn}\alpha_{gmi}}$ in (\ref{0061}) will increase. In addition, as the number of APs reduces, to maintain enough SINR in (\ref{0001}), the power that each AP allocates to a user will increase, then more interference will appear according to (\ref{0061}). Moreover, the mean-interference of proposed algorithms is less than the reference {two}. The main reason is that, in this paper, power allocation and user grouping are jointly optimized considering QoS constraints. To reduce the total transmit power and satisfy QoS requirements of different users as problem $\mathcal{P}1$ shows, the value of ${\sum\limits_{g=1}^{G}x_{gn}\sum\limits_{i=1}^{N}x_{gi}\sum\limits_{m=1}^{M}p_{mi}\beta_{mn}\alpha_{gmi}}$ in (\ref{0003b}) is reduced.
{Although interference is considered in the Gale-S strategy, the mean-interference of this strategy is high. The reason is that, as the users with lower target data rates need lower transmit power in general, the users with lower target data rates may be grouped into one group with less interference and the users with higher target data rates may be grouped into the other group. The interference among the desired signals of users in the group with higher target data rates is very serious.  Besides, it should be noted that, due to the inherent ability of ZF beamforming to null interference among the desired signals of users in each group, ZF beamforming is not shown in this figure.}

\begin{figure}[htb]
\setlength{\belowcaptionskip}{-0.5cm}
\setlength{\belowdisplayskip}{-1.2cm}
\centering
\subfigure[]{
\label{fig:Iu}
\includegraphics[height=0.18\textwidth]{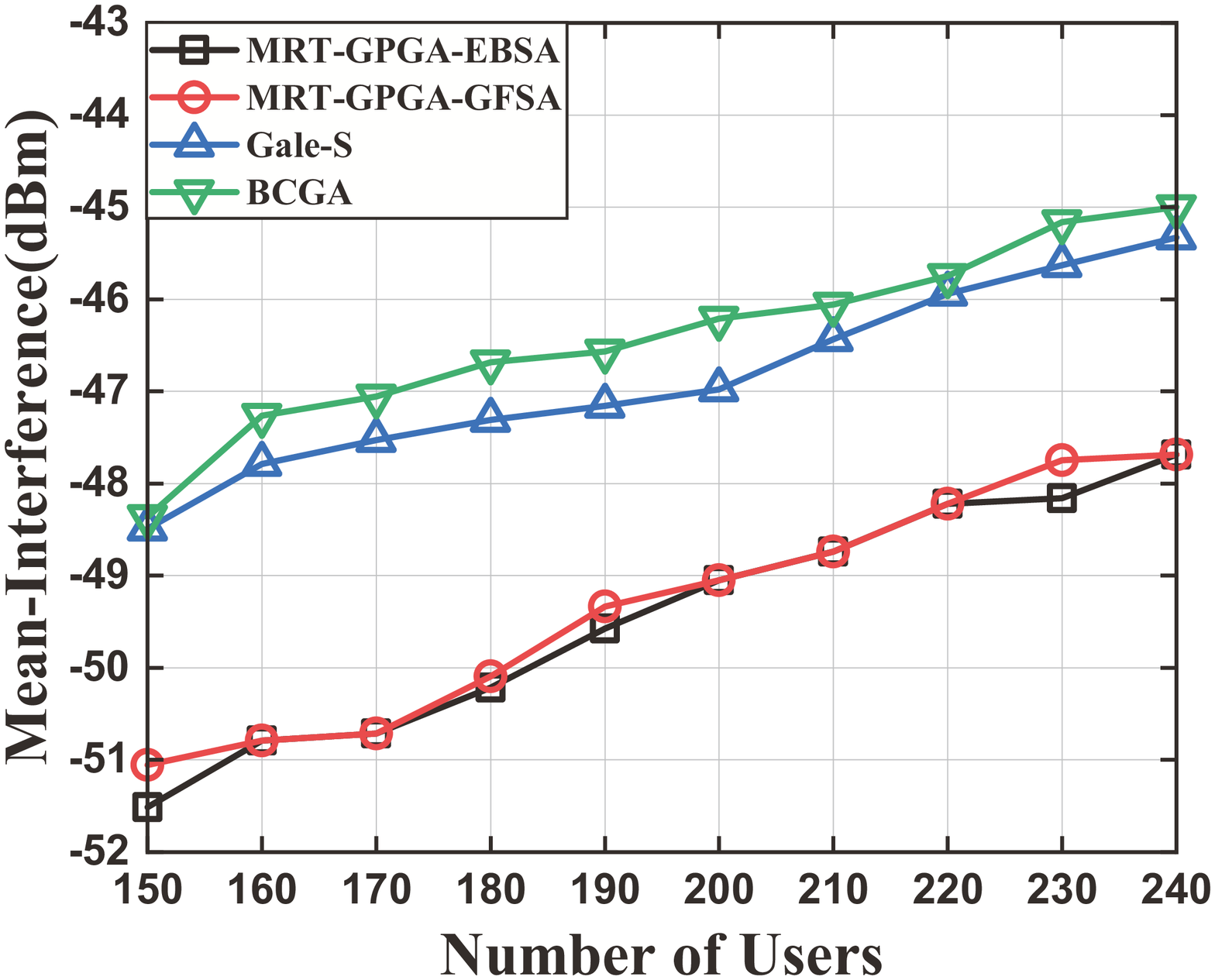}}
\subfigure[]{
\label{fig:IB}
\includegraphics[height=0.18\textwidth]{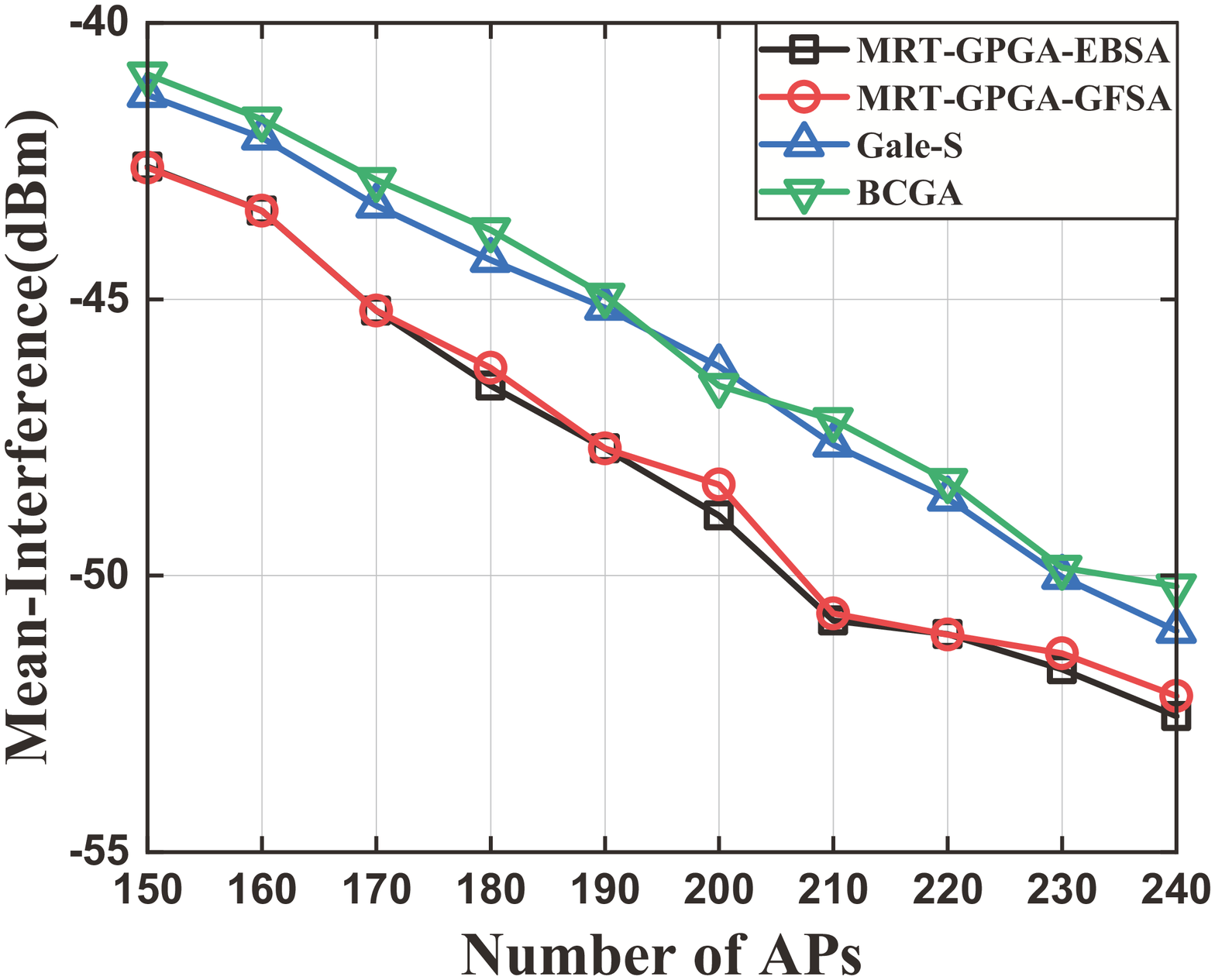}}
\centering
\caption{Performance comparison of different user grouping algorithms: a. Mean-interference vs. number of users; b. Mean-interference vs. Number of APs.}
\label{fig:UU}
\end{figure}

It is clear that interference has great impact on power allocation. In Fig. \ref{fig:Pu} and Fig. \ref{fig:PB}, we illustrate the total transmit power of {six} different user grouping algorithms with varied numbers of users and APs, respectively. The results show that all the {six} curves in Fig. \ref{fig:Pu} rise as the number of users increases and all the {six} curves in Fig. \ref{fig:PB} decrease as the number of APs increases. That is because that, as shown in Fig. \ref{fig:Iu} and Fig. \ref{fig:IB}, the mean-interference increases with the increase of users and the reduction of APs. Then more power is needed to maintain enough SINR. In addition, the total transmit power of the proposed user grouping algorithms is significantly lower than the reference {two}, which agrees with the results in Fig. \ref{fig:Iu} and Fig. \ref{fig:IB}. {Furthermore, with the proposed user grouping strategies, the transmit power consumption of ZF beamforming is lower than that of MRT beamforming. This result derives from the inherent ability of ZF beamforming to null interference among the desired signals of different users. }

\begin{figure}[htb]
\setlength{\belowcaptionskip}{-0.5cm}
\setlength{\belowdisplayskip}{-1.2cm}
\centering
\subfigure[]{
\label{fig:CT}
\includegraphics[height=0.18\textwidth]{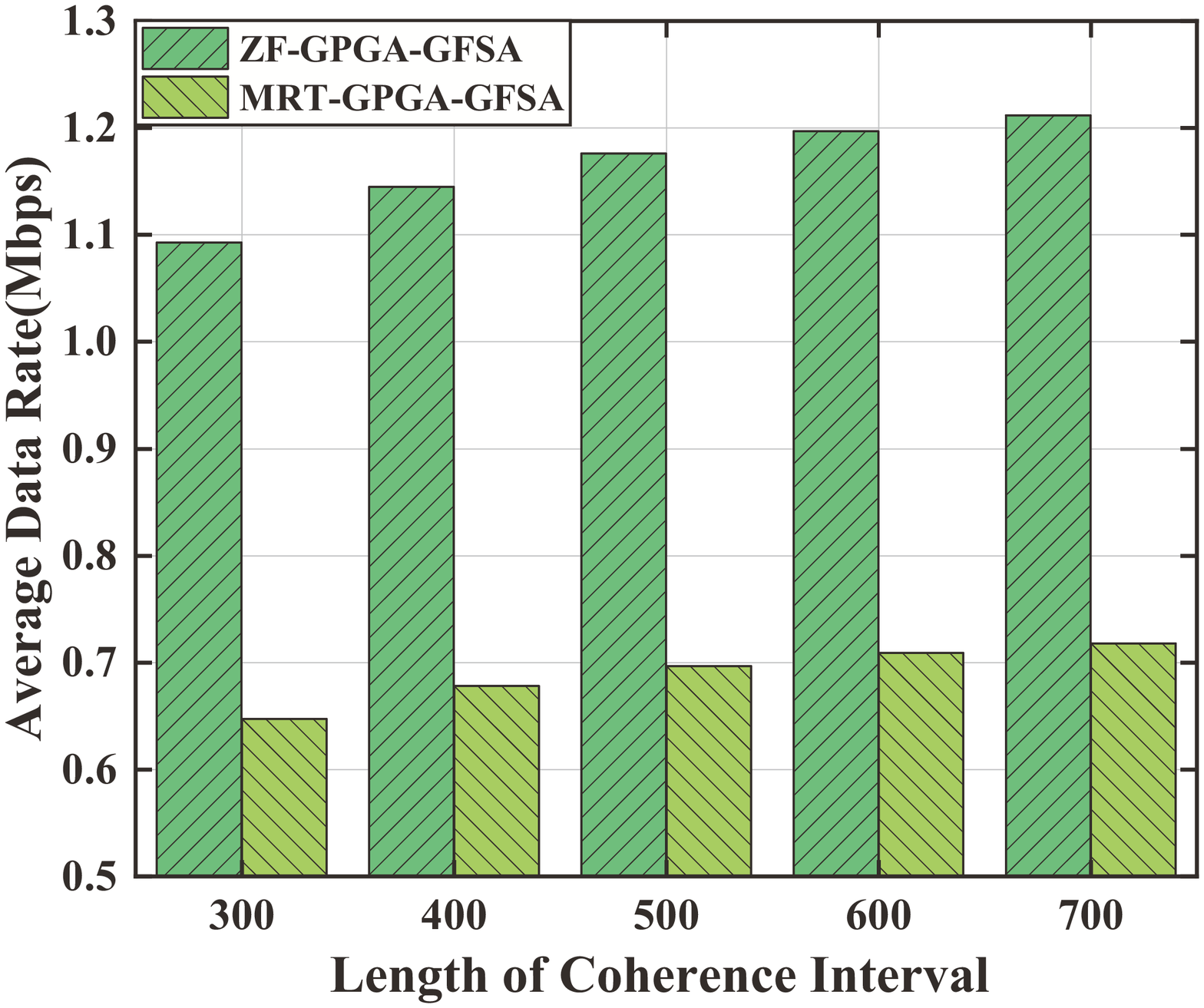}}
\subfigure[]{
\label{fig:GN}
\includegraphics[height=0.18\textwidth]{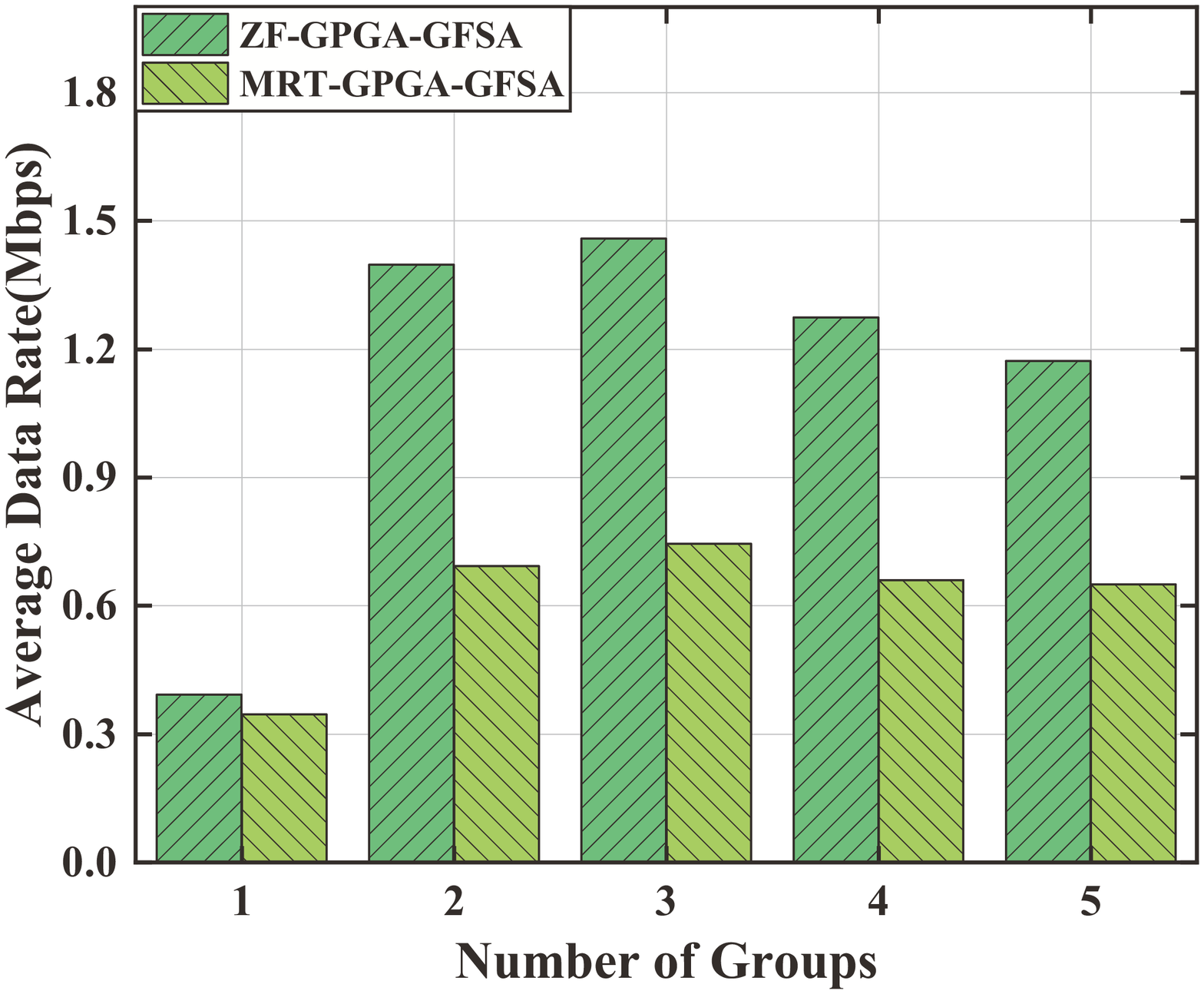}}
\centering
\caption{{Performance comparison under given total transmit power: a. Average data rate(Mbps) vs. Length of coherence interval(M=250, N=240, $P_t$=10W); b. Number of groups vs. Length of coherence interval(M=250, N=240, $P_t$=10W).}}
\label{fig:CTGN}
\end{figure}

{
\begin{figure*}[ht]
	\centering
	\includegraphics[scale=0.6]{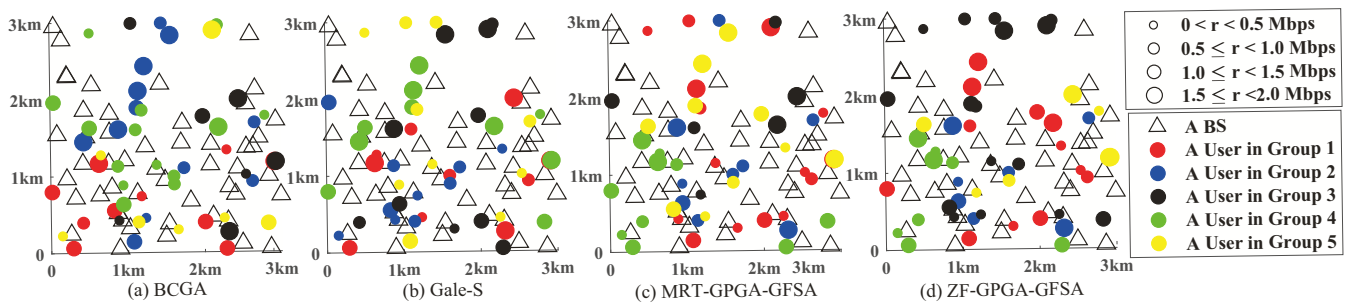}
	\caption{Distribution of users and APs in 3km $\times$ 3km rectangular area(50 users, 50 APs and 5 groups).}
	\label{fig:fb}
\end{figure*}}

\begin{figure*}[htb]
\setlength{\belowcaptionskip}{-0.5cm}
\setlength{\belowdisplayskip}{-1.2cm}
\centering
\subfigure[]{
\label{fig:BCGA}
\includegraphics[height=0.2\textwidth]{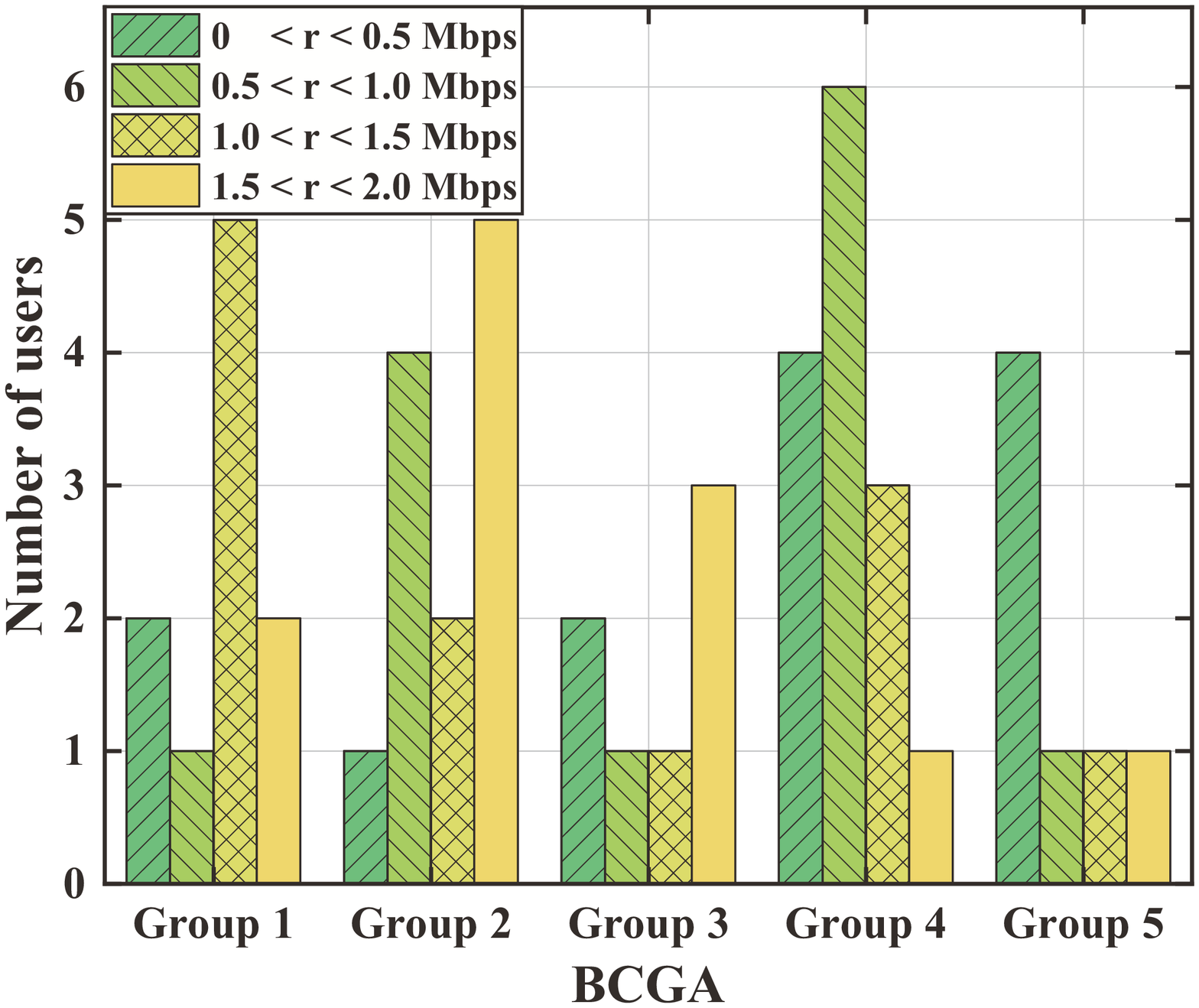}}
\subfigure[]{
\label{fig:Gale}
\includegraphics[height=0.2\textwidth]{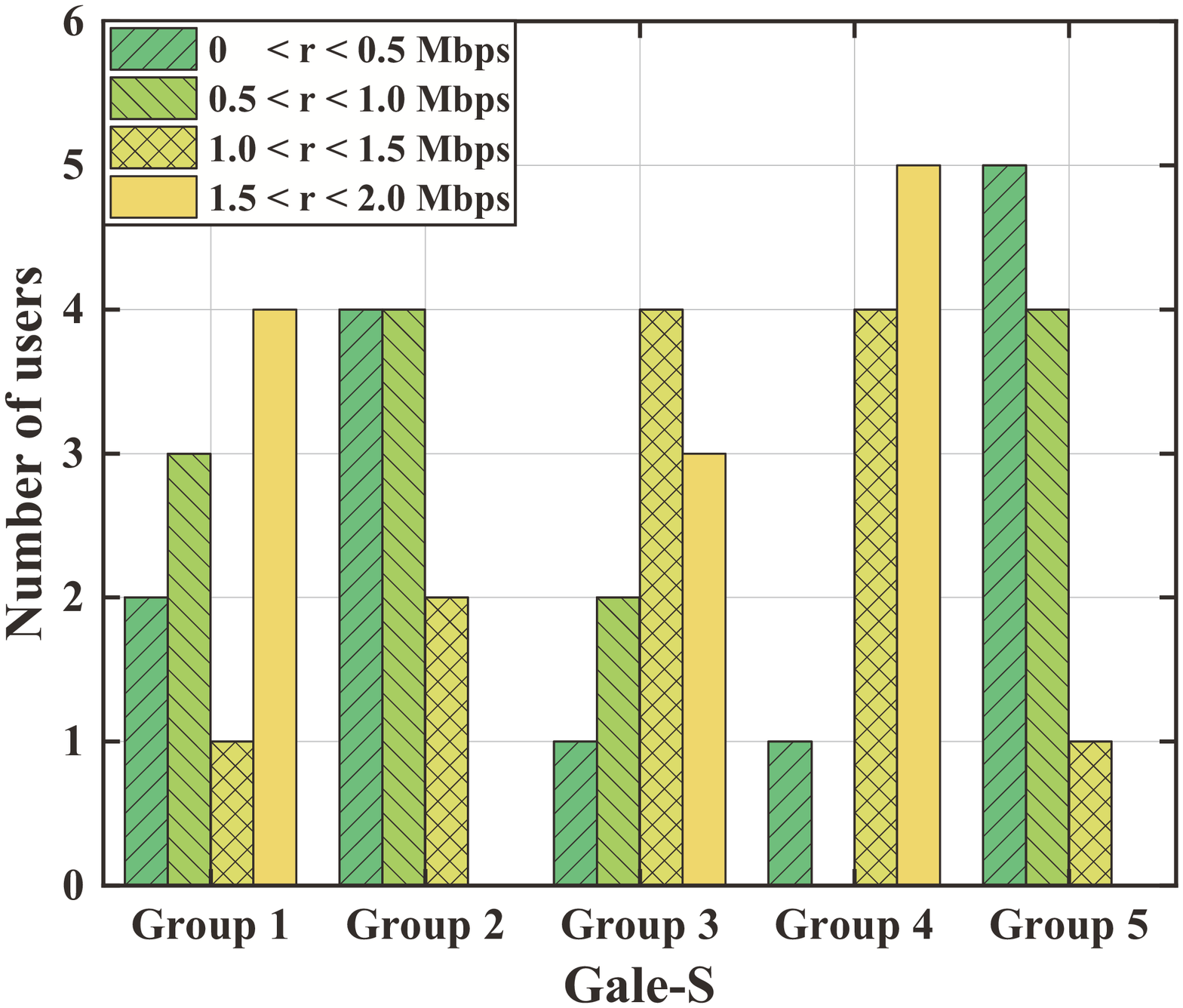}}
\subfigure[]{
\label{fig:MRT}
\includegraphics[height=0.2\textwidth]{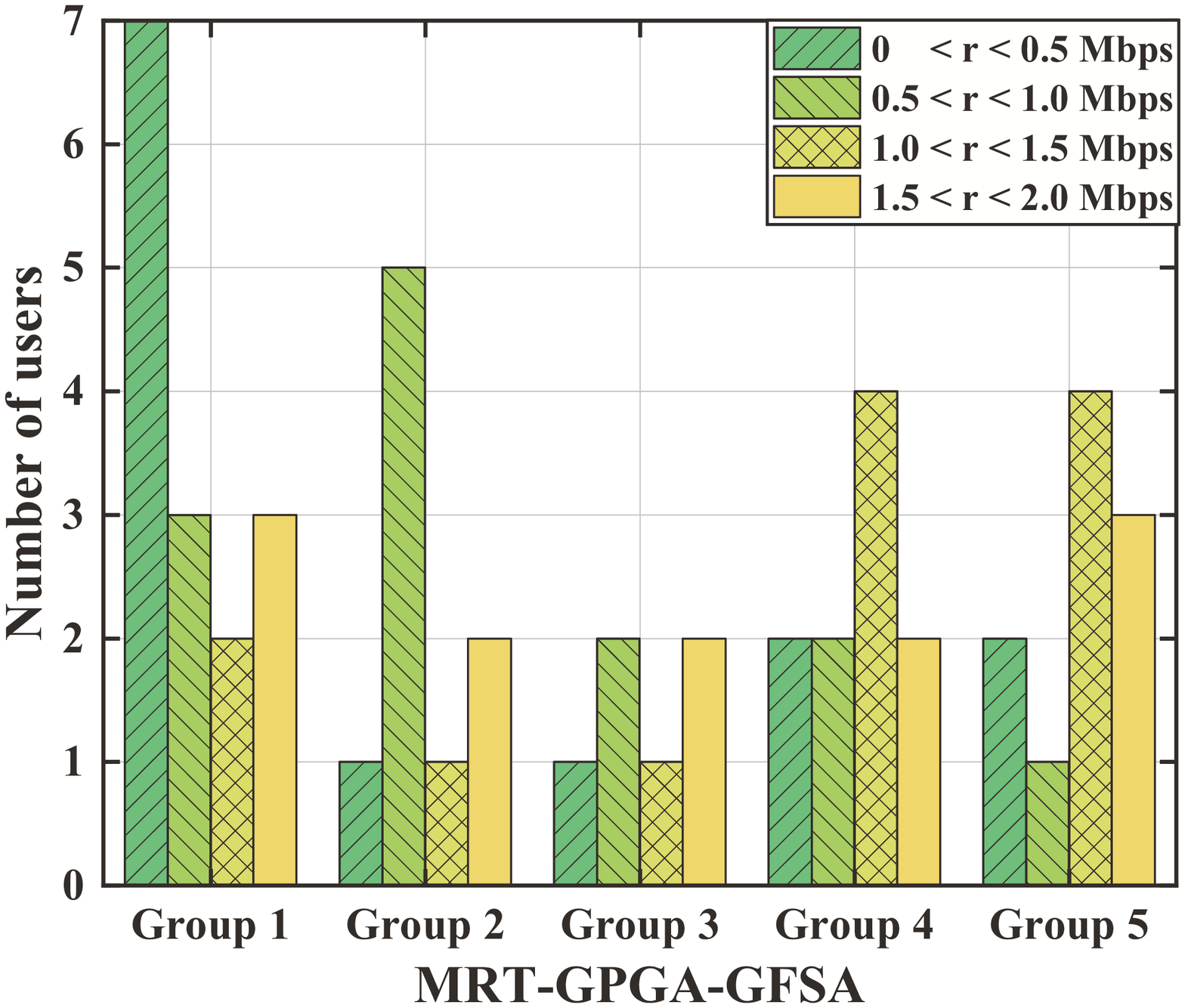}}
\subfigure[]{
\label{fig:ZF}
\includegraphics[height=0.2\textwidth]{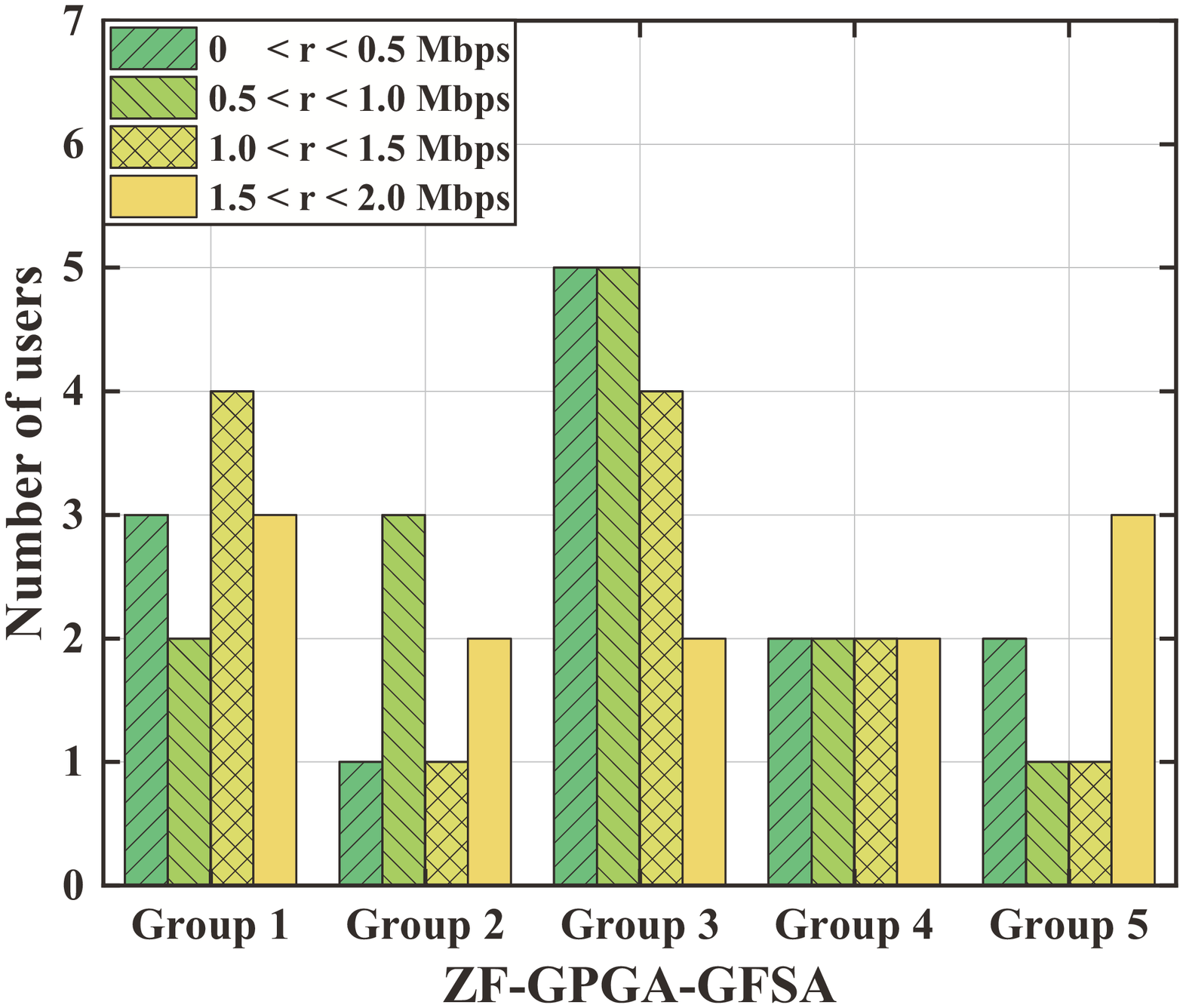}}
\centering
\caption{Number of users in each group and in each range of QoS requirements(50 Users, 50 APs and 5 Groups).}
\label{fig:fb2}
\end{figure*}
In Fig. \ref{fig:rr}, we show the total transmit power with different ranges of target data rate. The results show that the total transmit power increases with target data rate, and the proposed algorithms outperform the reference algorithm. The reason is that, target $\mathrm{SINR}$ of each user, i.e., $\gamma_n$, are explicit in constraints (\ref{0003b}) of problem $\mathcal{P}1$. {Furthermore, in Fig. {\ref{fig:BS}}, the total transmit power of the downlink cell-free massive MIMO system without user grouping (named ``Non-grouping''), is compared with that of the proposed system with user grouping. We can see that the column without grouping is higher than the other columns. This result shows that, after user grouping, the minimal transmit power {required for satisfying} all users' QoS requirements is reduced. An important reason is that, after user grouping, the length of pilot sequences, i.e., $\tau_g$, can be reduced without reducing the accuracy of channel estimation as shown in Fig. \ref{fig:51}. Then the number of symbols for data transmission within each coherent interval, i.e., $\tau_c-\sum_{g=1}^{G}\tau_g$, can be increased.}

{ In Fig. \ref{fig:CTGN}, under given total transmit power, the average data rate is evaluated with different lengths of coherence interval and group numbers. In Fig. \ref{fig:CT}, we vary the length of coherence interval to show the average data rate with different beamforming methods. As the length of coherence interval increases, the symbols for data transmission will increase, and the length of pilot sequence is unchanged. Therefore, the average data rate will increase. Furthermore, the average data rate of users with ZF beamforming is higher than that of users with conjugate beamforming, the reason is that the interference among the desired signals of different users can be cancelled by ZF beamforming. In order to further demonstrate the advantages of user grouping in cell-free massive MIMO systems, in Fig. \ref{fig:GN}, we show the average data rate vary with the number of groups. We can see that the average data rate can be improved by user grouping. In general, by beamforming among the antennas of many APs, more gains from spatial diversity can be obtained. After user grouping, the number of users sharing the same time-frequency resource will be reduced, and the utilization efficiency of spatial diversity is reduced. However, the pilot overheads will be greatly reduced by user grouping as shown in section V.B. There is a tradeoff between the utilization efficiency of spatial diversity and the pilot overheads. Therefore, in Fig. \ref{fig:GN}, the average data rate will increase from $G=1$ to $G=3$, and descend when the number of groups is greater than three.}

The number of users sharing the same time-slot in cell-free massive MIMO systems is much larger than traditional communication systems where radio resources of different users are usually orthogonal. {In general, the users with higher QoS requirements require more transmit power from APs to guarantee a certain SINR. More transmit power will lead to more severe interference to the desired signals of the other users in the same group. If the users with high QoS requirements be assigned into the same group, the interference among the signals of users in the same group will be too serious to be eliminated and the power consumption will be unbearable. Therefore, to alleviate the serious interference from the desired signals of the other users in the same group}, users with high QoS requirements should be avoided to be assigned into the same group. In Fig. \ref{fig:fb}, we show the distribution of users and APs with different user grouping algorithms, where users and APs are represented by dots and triangles, respectively. The number of APs is 50. There are 50 users which are assigned into 5 groups, and the dots in the same colour represent the users in the same group. The size of each dot reflects its QoS requirement as shown in the legend of Fig. \ref{fig:fb}. In addition, to analyze the distribution of users and APs in Fig. \ref{fig:fb} more intuitively, we also show the number of users in each group and in each range of QoS requirements in Fig. \ref{fig:fb2}. {The results show that there are 5 users whose target data rates are greater than $1.5$Mbps in group $2$ with user grouping algorithm BCGA, and there are 5 users whose target data rates are greater than $1.5$Mbps in group $4$ with user grouping algorithm Gale-S, which will bring serious interference from the desired signals of the other users} in the same group to the users in this group. By contrast, users with high target data rates are separated into different groups in the proposed algorithms. This also explains why the proposed algorithms outperform the reference algorithm in terms of transmit power, {interference from the desired signals of the other users in the same group} as shown in Fig. \ref{fig:Pu}, Fig. \ref{fig:PB}, Fig. \ref{fig:Iu} and Fig. \ref{fig:IB}. Moreover, the number of users in each group is no more than $16$ as Fig. \ref{fig:fb2} shows. In other words, length of pilot sequence can be effectively reduced by user grouping.

\section{Conclusion}

In this paper, we study the joint optimization problem of power allocation and user grouping to minimize the total transmit power in cell-free massive MIMO systems. We decompose this problem into a primal problem: power allocation problem and a master problem: user grouping problem, where the power allocation problem is proved to be convex. We analyze and relax these two problems by GBD method. Then an algorithm based on GBD method is proposed to solve the joint optimization problem by iteratively solving these two problems and reduce the gap between the upper bound and lower bound of the original problem. Moreover, the relaxed master user grouping problem is converted into a problem of searching for some special negative loops in a graph composed of users based on graph theory. An algorithm extended from Bellman-Ford algorithm as well as a fast greedy suboptimal algorithm is proposed to search for these negative loops.

Although the complexity of channel estimation and decoding can be reduced by user grouping, there still remain some challenges in research on cell-free massive MIMO as the number of users increases. For instance, each AP needs to know the transmitted symbols of all users after user grouping, so the limited fronthaul is still one of the bottleneck in cell-free massive MIMO systems as the number of users increases. An effective method for fronthaul reduction is to reducing the number of APs connected with each user (AP grouping). However, the benefit of spatial diversity will also decrease after AP grouping. To serve more users in cell-free massive MIMO systems with limited APs and limited fronthaul, there still remain many works to do.

\bibliographystyle{IEEEtran}
\bibliography{bibtex}

\end{document}